# Six-sigma Quality Management of Additive Manufacturing

Hui Yang[1], Prahalad Rao[2], Timothy Simpson[1,4], Yan Lu[3], Paul Witherell[3], Abdalla R. Nassar[4], Edward Reutzel[4], and Soundar Kumara[1]

[1] Harold and Inge Marcus Department of Industrial and Manufacturing Engineering
The Pennsylvania State University, University Park, PA, USA

[2]Department of Mechanical & Materials Engineering
University of Nebraska–Lincoln, Lincoln, NE, USA

[3]National Institute of Standards and Technology
Gaithersburg, MD, USA

[4]Center for Innovative Materials Processing 3D (CIMP-3D)
The Pennsylvania State University, University Park, PA, USA

## Abstract

Quality is a key determinant in deploying new processes, products or services, and influences the adoption of emerging manufacturing technologies. The advent of additive manufacturing (AM) as a manufacturing process has the potential to revolutionize a host of enterprise-related functions from production to supply chain. The unprecedented level of design flexibility and expanded functionality offered by AM, coupled with greatly reduced lead times, can potentially pave the way for mass customization. However, widespread application of AM is currently hampered by technical challenges in process repeatability and quality management. The breakthrough effect of Six Sigma has been demonstrated in traditional manufacturing industries (e.g., semiconductor and automotive industries) in the context of quality planning, control, and improvement through the intensive use of data, statistics and optimization. Six sigma entails a data-driven DMAIC methodology of five steps - **D**efine**, M**easure, **A**nalyze, **I**mprove, and **C**ontrol. Notwithstanding the sustained successes of Six-Sigma knowledge body in a variety of established industries ranging from manufacturing, healthcare, logistics and beyond, there is a dearth of concentrated application of Six-Sigma quality management approaches in the context of AM. In this paper, we propose to design, develop, and implement the new DMAIC methodology for Six-Sigma quality management of AM. First, we define the specific quality challenges arising from AM layer-wise fabrication and mass customization (even one-of-a-kind production). Second, we present a review of AM metrology and sensing techniques, from materials through design, process, environment, to post-build inspection. Third, we contextualize a framework for realizing the full potential of data from AM systems, and emphasize the need for analytical methods and tools. We propose and delineate the utility of new data-driven analytical methods, including deep learning, machine learning, and network science, to characterize and model the interrelationships between engineering design, machine setting, process variability and final build quality. Fourth, we present the methodologies of ontology analytics, design of experiments (DOE) and simulation analysis for AM system improvements. In closing, new process control approaches are discussed to optimize the action plans, once an anomaly is detected, with specific consideration of lead time and energy consumption. We posit that this work will catalyze more in-depth investigations and multi-disciplinary research efforts to accelerate the application of Six-Sigma quality management in AM.

# I. Introduction

Additive manufacturing (AM) is a collective term for processes in which a product is made by layer-upon-layer deposition of materials. The advent of commercial AM systems has enabled the fabrication of parts with complex geometry directly from computer-aided design (CAD) models with minimal intervening steps. Until recently, AM parts were primarily restricted to prototype-demonstrator roles; the viability of AM parts have now evolved to the extent that they are used in production and final assemblies. The choice of materials ranges from metals, composites, polymers to ceramics. AM provides significant advantages over traditional subtractive (machining) and formative (casting, welding, molding) manufacturing processes, such as eliminating specialized tooling costs, reducing material waste and life-cycle costs, enabling the creation of intricate and free-form geometries, and expanding product functionality for a variety of industrial applications.

However, technical challenges in quality management hamper widespread adoption of AM technology in the industry. For example, microstructure and mechanical properties of AM builds are influenced by complex, hard to model, process phenomena (e.g., thermal effects, residual stresses). These intricate process interactions, in turn, can lead to hidden internal defects that deteriorate the quality of the part. As a result, the rejection rate of AM parts is high, particularly when considering one-of-a-kind production. In real-world case studies, it is not uncommon that parts that are built simultaneously with the same CAD model in the same commercial AM machine may yield different quality results (see Figure 1). The high rejection rate of AM builds and associated costs significantly hinder the wider exploitation of AM capabilities, beyond the current rapid prototyping status quo.

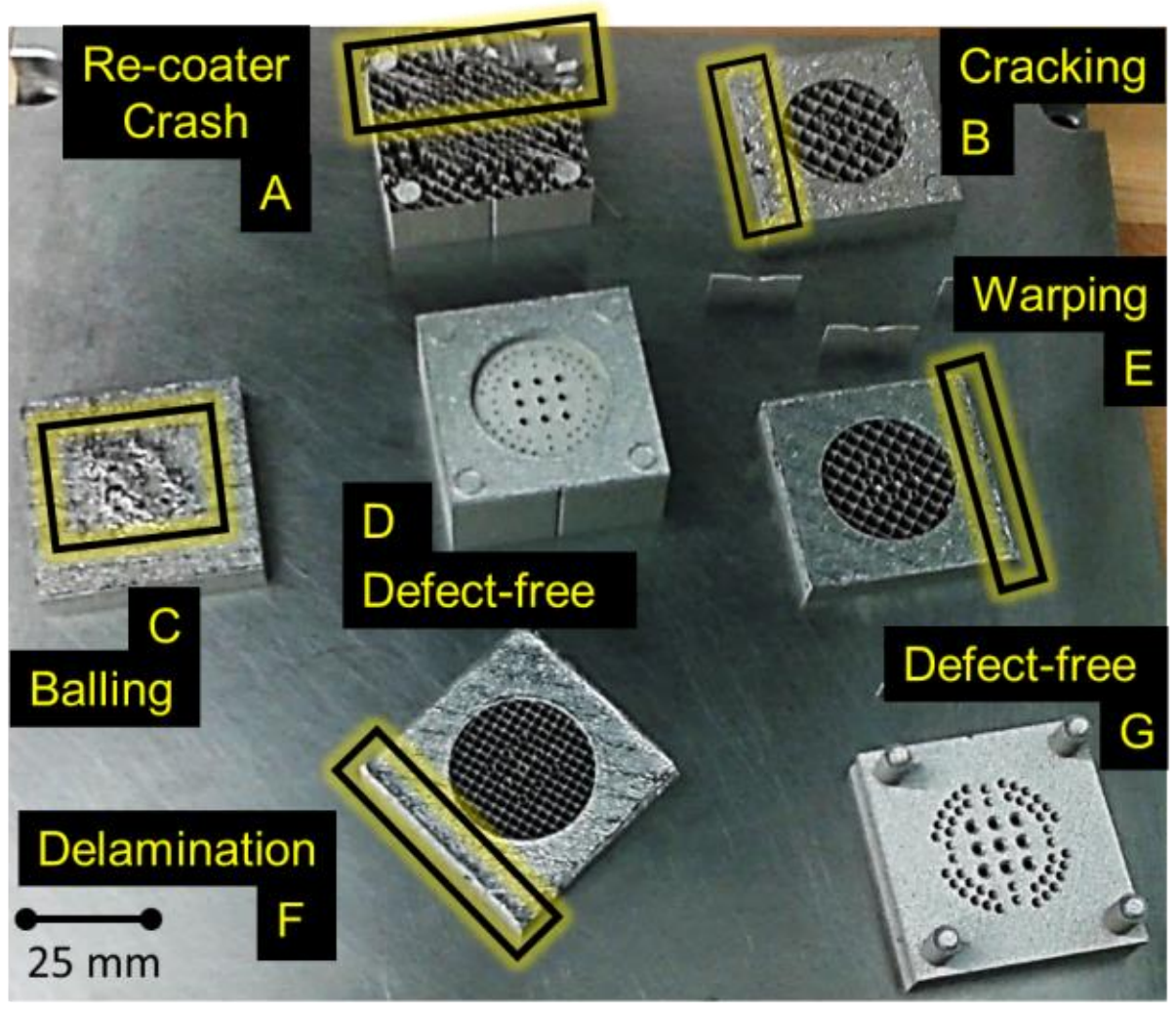

**Figure 1. Seven stainless steel parts built on a commercial AM system. The parts only differ in their orientation, with all other process conditions identical [1]**

Six Sigma (6S) is a widely used practice in traditional manufacturing industries (e.g., semiconductor and automotive industries) for quality planning, quality assurance (QA), quality control (QC), and continuous improvements with the extensive use of data, statistics and optimization [2, 3]. As shown in Figure 2, Six Sigma entails a data-driven **D**efine**, M**easure, **A**nalyze, **I**mprove, and **C**ontrol (DMAIC) methodology consisting of five phases: **1) Define**: outline the quality challenges based on the customer requirements; **2) Measure**: collect data about key process variables from the manufacturing systems; **3) Analyze**: extract useful information pertinent to defect-causing factors; **4) Improve**: design solutions and methods to improve the manufacturing system; **5) Control**: develop process management plans and optimal control policies when the manufacturing system is out of control. The goal of Six Sigma techniques is to identify and remove root causes of defects, and further improve the quality of final products. The success of Six Sigma can be seen through Motorola's application of its philosophies. In 1978, the company had a net income of $2.3 billion. By 1988, the net income had increased to $8.3 billion; this is roughly a 260% increase. Similarly, General Electric saw massive successes with their own 6S program and achieved $4 billion in savings per year. The list goes on with other notable examples, including Toyota, Ford, Polaroid, General Motors, and many more.

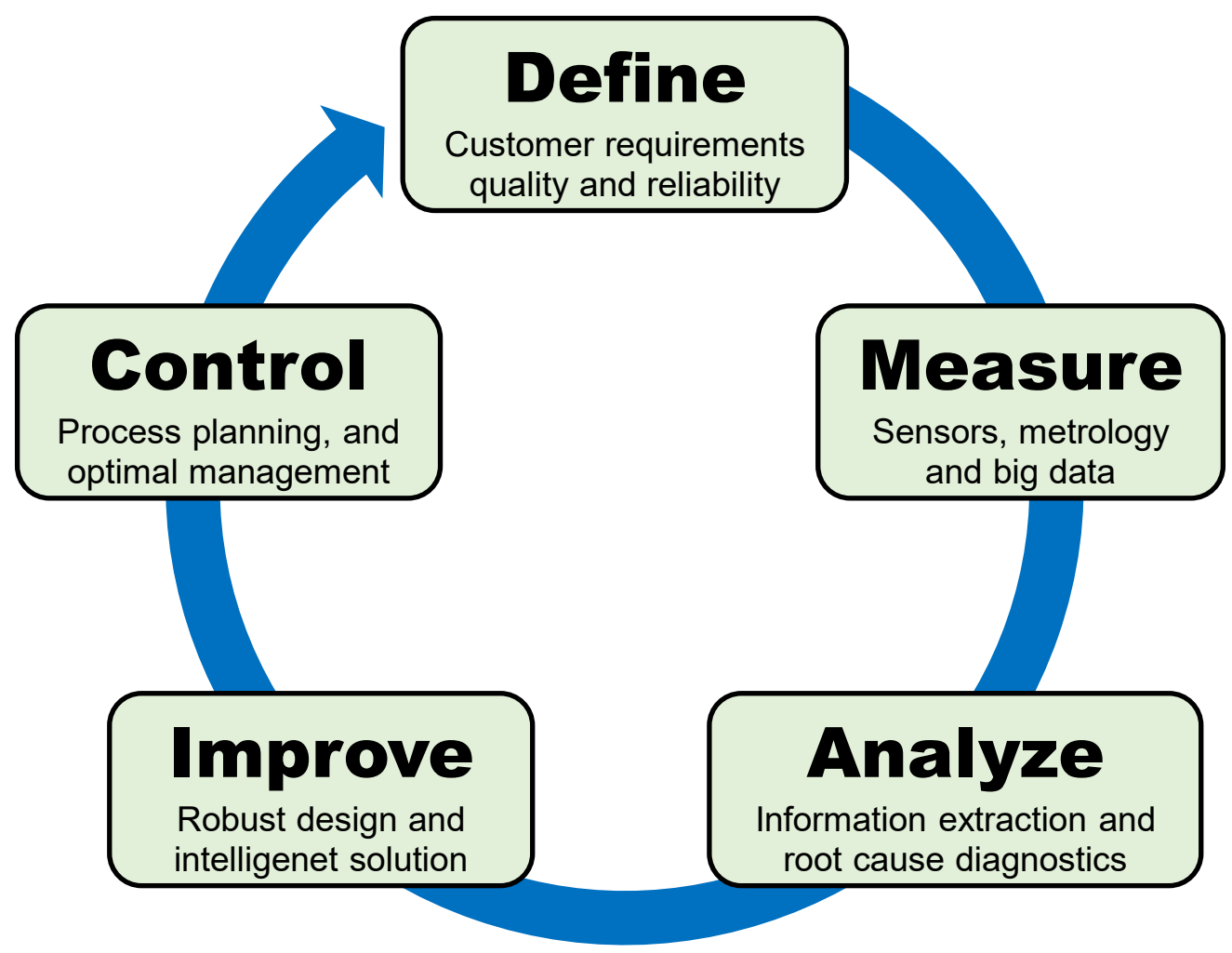

**Figure 2. The DMAIC methodology for Six-Sigma quality management**

Although Six Sigma has achieved significant success in a host of domains ranging from manufacturing, healthcare, and logistics, more research needs to be done to initiate the practice of Six-Sigma quality management in the specific context of AM. In this paper, we propose to design, develop, and implement the new DMAIC methodology for Six-Sigma quality management of AM. First, we define the specific quality challenges arising from AM layer-wise fabrication and mass customization (even one-of-a-kind production). Second, we present a review of AM metrology and sensing techniques, from materials through design, process, environment, to post-build inspection. Third, realizing the full potential of AM sensing data depends to a great extent on the availability of analytical methods and tools. Accordingly, we propose and develop new data-driven analytical methods, including deep learning, machine learning, and network science, to characterize and model the interrelationships between engineering design, machine setting, process variability, and final build quality. Fourth, we present the methodologies of ontology modeling, design of experiments (DOE) and simulation analysis for continuous quality improvements. In the end, new control approaches are discussed to optimize the action plans, once an anomaly is detected, with specific considerations of lead time and energy consumption. The goal of this paper is to catalyze more in-depth investigations and multi-disciplinary research efforts to lay the foundation of a new scientific basis of Six-Sigma quality management for AM.

The rest of the paper is organized as follows: Section II discusses specific quality challenges arising from unique AM characteristics such as mass customization (even one-of-a-kind), low-volume production, multi-layer part fabrication, and sequential manufacturing. Section III reviews the development of advanced sensing and measurement systems to increase information visibility for AM quality management. Then, we present AM data analytics in Section IV. Continuous quality improvements for AM are discussed in Section V, and Section VI presents the sequential optimization of layerwise control strategies for AM. Section VII discusses the Six-Sigma quality management for AM and concludes this paper.

## II. Define Quality Challenges

AM's capability to build objects from the ground up stimulates the imagination, causing one to envision a broader range of possibilities during design. Nonetheless, AM faces a broad range of quality challenges that hamper the wider adoption of AM in the industry. The urgent need to produce complex builds in low volume and high mix, combined with rapid advancements in AM technology, poses significant challenges on current paradigms for AM quality management. As such, new standards are being developed for material and process qualification and part certification [4, 5], countless experiments and modeling/simulation studies are being conducted to gain insights into the complex physics of AM processes [6-8], new in-situ sensing capabilities and process monitoring strategies are developed for process control [9-13], and efforts are underway to capture, store, manage, and assure pedigreed data for QA/QC of AM parts [14, 15]. In spite of these advances, repeatability and reliability issues seen in many metal AM processes (e.g., laser powder bed fusion (LPBF), directed energy deposition (DED)) unfortunately exacerbate these challenges, particularly when trying to produce end-use parts for critical applications and highly regulated industries (e.g., aerospace, medical) [16-19].

## A. Quality management for high-volume-low-mix production

It is well known that "quality is inversely proportional to variability" [20, 21]. As shown in Figure 3, the measure of "variability" often refers to the scenario of high-volume-low-mix production in the context of mass manufacturing. In other words, if there are a large quantity of parts produced from the manufacturing systems, then it will be a logical step to characterize and measure the process variability and repeatability. The variability can be due to random or assignable causes in the manufacturing process. If the quality variations are solely because of random factors (i.e., non-assignable causes, not identifiable), then the distribution should be normal. However, if there are assignable causes, then statistical control charts are often used to monitor the process and detect when and how the process performance is affected. As such, the process can be stopped to look for assignable causes and eliminate them to resume the normal production. Quality improvement involves a series of managerial, operational, and engineering activities to reduce the variability in the process. Specifically, statistical design of experiments is utilized to realize a robust process by studying the effects of controllable settings under the uncertainty of uncontrollable factors, also called "robust parameter design" [22].

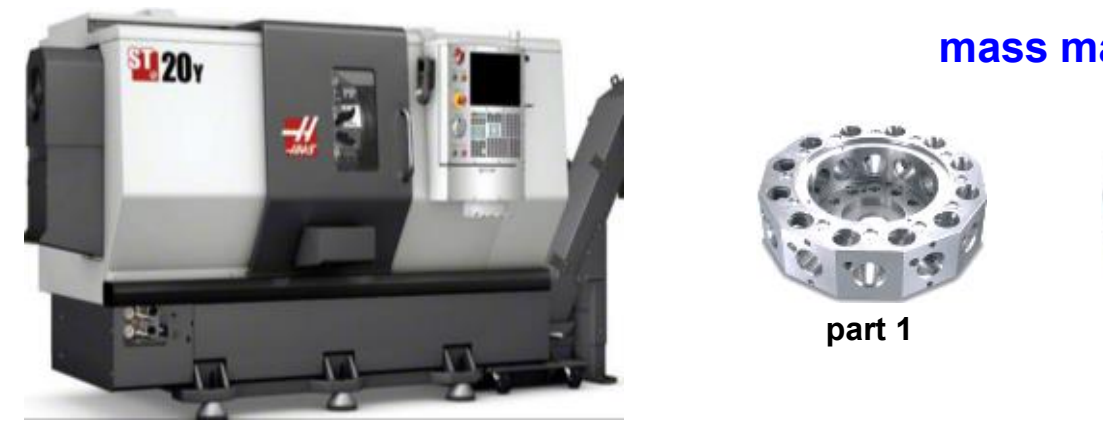

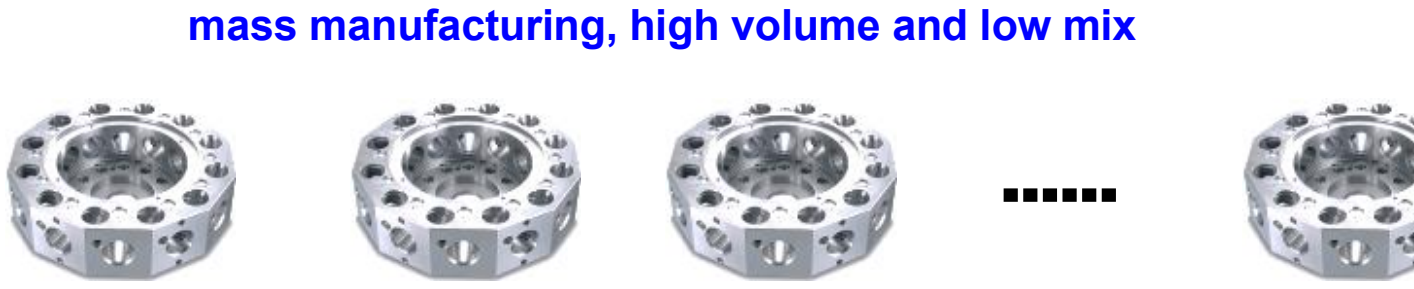

**Figure 3. The high-volume-low-mix production scheme in mass manufacturing**

As a result, the Six-Sigma program emerged to meet the needs of mass manufacturing in automotive and semiconductor industries, and has achieved enormous successes in the past century. As shown in Figure 4, the Six-Sigma program utilizes the DMAIC methodology for the reduction of process variability to the level that failures and defects are extremely unlikely. If the $3\sigma$ limits overlap with product specification limits, then the probability for a part falling outside the $\mu \pm 3\sigma$ limit is 0.27%, which means that the number of defective parts per million (PPM) is about 2,700. For the $\mu \pm 6\sigma$ limit, the probability will be 0.0000002%, which means that the PPM is 0.002 (i.e., extremely unlikely). In the Six-Sigma scenario, if a finished product has 100 components and each component must be non-defective for the product to be non-defective, then the probability of the product to be non-defective is $(0.999999998)^{100} \approx 1.0$. The Six-Sigma concepts (e.g., design for Six-Sigma, lean production, variation reduction) have been widely used to improve the capability of many business processes nowadays. The development of Six-Sigma program has gone through three phases as follows:

- Phase I: address process monitoring, defect elimination and variability reduction
- Phase II: reduce total production cost and increase system performance
- Phase III: emphasize the value creation to business organizations

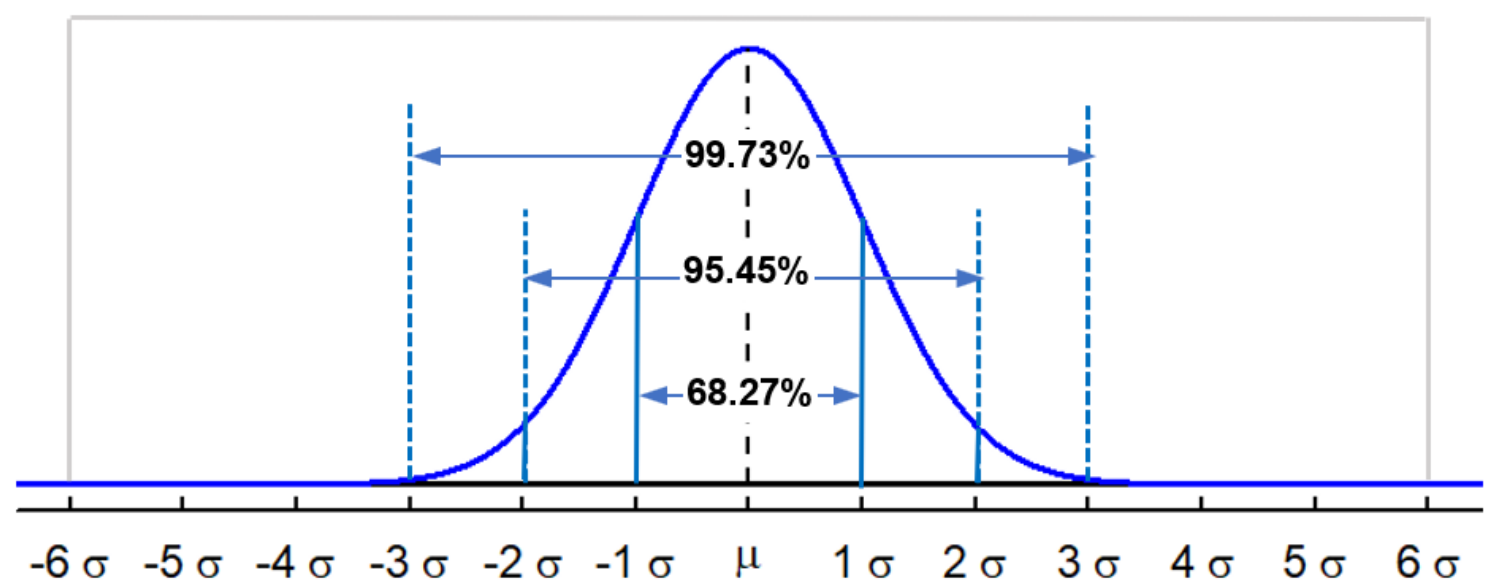

| Specification limit | % inside specs | PPM defective |
|---|---|---|
| $\mu \pm 1\sigma$ | 68.27 | 317,300 |
| $\mu \pm 2\sigma$ | 95.45 | 45,500 |
| $\mu \pm 3\sigma$ | 99.73 | 2,700 |
| $\mu \pm 4\sigma$ | 99.9937 | 63 |
| $\mu \pm 5\sigma$ | 99.999943 | 0.57 |
| $\mu \pm 6\sigma$ | 99.9999998 | 0.002 |

**Figure 4. The area under the normal curve and the proportion of defectives produced**

However, additive manufacturing moves towards the high level of customization by enabling low-volume-high-mix production (even one-of-a-kind production) directly from the digital designs from the customers, resulting in "economies of one" [23]. The large quantity of parts produced from the same design are not available any more, as in the traditional paradigm of mass manufacturing, to establish and measure process variability. Therefore, the Six-Sigma practice from mass manufacturing tends to be limited in the ability to be generally applicable to AM. There is an urgent need to push forward the next phase of Six-Sigma program for AM. As shown in Figure 5, AM presents new QA/QC challenges: mass customization, low-volume production, and layer-to-layer variations in part geometry. In particular, because of the customized design and layer-by-layer fabrication in AM, it is not uncommon that each layer is different from another layer in terms of part geometry. Hence, it is difficult to characterize and measure the process variability and repeatability from one layer to another or from one build to the next.

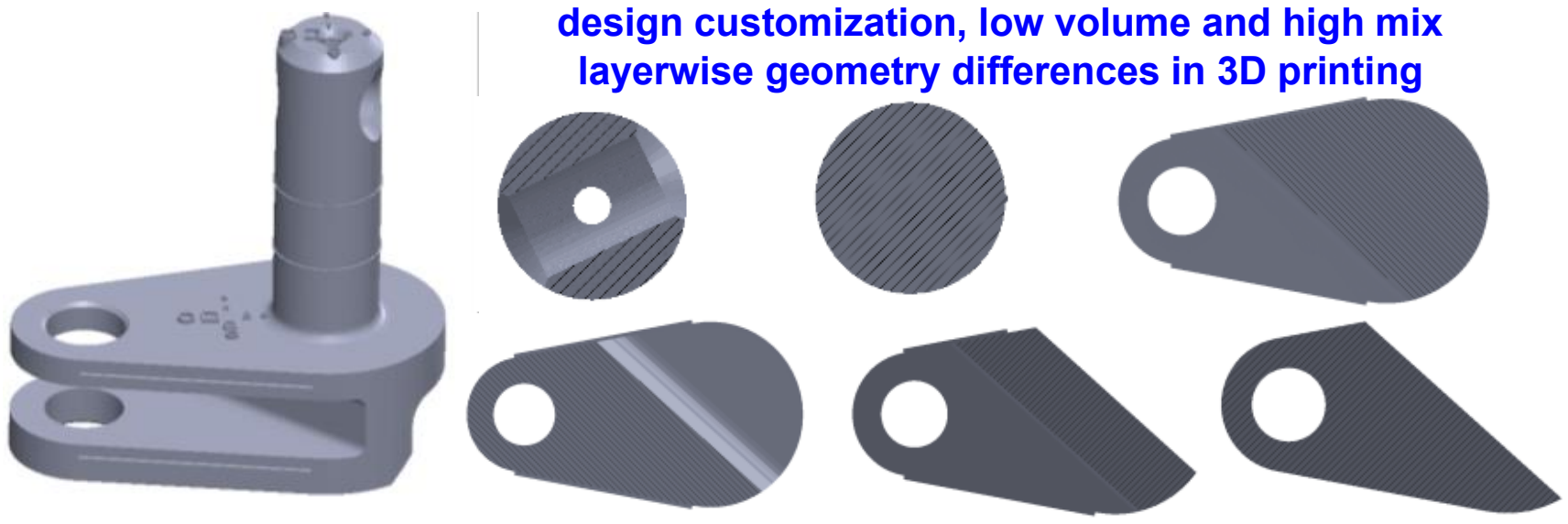

**Figure 5. The low-volume-high-mix production scheme for a customized design with layer-to-layer variations in part geometry in 3D printing**

## B. Multi-layer and sequential manufacturing process

The layer-by-layer approach to AM brings significant challenges for QA/QC. Many AM processes use the raw materials of metal powders, where particle sizes and shapes vary between batches. Also, a laser or electron beam is utilized as the heating source in LPBF and DED. Slight variations in the intensity and diameter of the beam contribute to the issue of repeatability both between different machines and between the same machine at different locations on the build plate. Thus, every parameter that affects the end result of the process must be tailored to the materials used [24]. Furthermore, an AM system can utilize different layer thicknesses when manufacturing parts. A 2cm high object that uses a layer thickness of $100\mu m$ will require 200 layers. Should the layer thickness be 50 $\mu m$, the number of layers would be 400. Each of these layers has the opportunity for failure. Even if a single layer has a small probability of having a defect, the overall build will have a high probability of having at least one defect. To illustrate the effects and challenges of multi-layer fabrication, consider the following example:

- If the probability to contain defects is 0.0114 in a layer, then what is the probability for this layer to be non-defective?

$$1 - 0.0114 = 98.86\%$$

- For an AM build with 100 layers, what is the probability to have no defects?

$$(1 - 0.0114)^{100} = 31.77\%$$

- For an AM build with 100 layers, what is the probability of having at least a defect?

$$1 - (1 - 0.0114)^{100} = 68.23\%$$

- If the probability of a build to contain defects is specified to be less than 10%, then what should be the probability for a layer to have defects?

$$1 - (1 - x)^{100} = 10\% \Rightarrow x = 0.0011$$

It is worth noting that this example assumes that each layer is independent of each other. However, AM is highly correlated from one layer to another layer. In other words, the defects in one layer can be corrected during processing of the subsequent layer, or can negatively impact the next layer and all the subsequent layers. This is analogous to the multi-stage assembly line in the traditional manufacturing paradigm. In the automotive industry, a car body assembly often involves a sequence of assembly operations. The variations in one assembly step can potentially introduce a stream of variations in the following steps [25]. However, the physics of multi-stage assembly operations are different from multi-layer AM with LPBF in each layer. A Six-Sigma program for multi-stage manufacturing systems typically analyzes the current state of a process and then incrementally improve system performance with statistical methods and tools.

Establishing a Six-Sigma paradigm for AM calls upon new innovations to tackle these emerging quality challenges, including mass customization, low-volume production, lay-to-layer variations, and multi-layer manufacturing process, which are unique when moving from traditional mass production to the new paradigm of AM. "**Measure**" requires the design and development of new sensor technologies for materials, processes, and post-build inspections at different stages of AM. "**Analyze**" should be able to handle and connect the big data that are generated during the AM product lifecycle. "**Improve**" calls upon a better understanding of the process physics and an ontological knowledge of the underlying phenomena through statistical design of experiments on physical machines, AM processes, and/or computer experiments on simulation models. "**Control**" should consider the sequential decision-making problem for multi-layer fabrication process in AM, and further address the multi-objective optimization of AM, e.g., minimizing total cost (e.g., energy or time) consumed in the LPBF process and maximizing the quality of final parts. The new scientific basis of six-sigma quality management will impact production-scale viability of AM, and enable wider exploitation of AM capabilities beyond the current rapid prototyping status quo.

## III. Measure AM

In the DMAIC approach, the measure step is aimed at collecting data from key variables during the AM process such as (1) process input variables (e.g., characteristics of metal powders, design parameters); (2) in-situ variables (e.g., machine settings, layerwsie imaging, thermal maps); and (3) process output variables (e.g., post-build CT scans). Modern manufacturing industries have invested in advanced sensing and measurement systems to cope with high levels of complexity in AM and increase the information visibility about key variables from raw materials, manufacturing process to final products. As aforementioned in Section II, the low-volume-high-mix production presents specific challenges on AM quality management. With rich data readily available from the step of "measure AM", this provides an opportunity for the "analyze" step to develop an in-depth understanding of current state and performance of AM process. Here, data could be collected online (i.e., in the layer-by-layer fabrication process), or offline (i.e., pre-build material characterization or post-build CT scan). Offline measurements allow for the inspection of quality but are limited in the ability to help in-process corrections or repairs, because the defects are often embedded within the build already. Online sensing captures the dynamics of process-machine interactions and offers a higher level of flexibility for on-the-fly control actions. The data collected in the "measure" step can be visualized in different ways to provide comprehensible information about the AM process, e.g., image stacks, 3D point clouds, histograms, network representation, Fourier and wavelet transformations. An effective visualization further helps the "analyze" step to estimate and extract salient features about the process variability or product defects.

## A. Pre-build measurement and characterization

As shown in Figure 6, metal powders are used as the input to the LPBF (and many DED) AM machines. Material qualification is indispensable to avoid the scenario of "garbage in, garbage out". Standard powder characterization techniques include x-ray photoelectron spectroscopy, sieve analysis, inert gas fusion, scanning electron microscopy, laser light diffraction, differential thermal analysis. These techniques allow the characterization of powders in three main aspects: particle morphology and distribution (e.g., the shape, surface roughness, or size), powder chemistry (i.e., elemental composition), and powder microstructure (e.g., porosity, rheology), see [26] for a review of AM powder characterization. The standard practices for sampling metal powders are provided by standards organizations such as ASTM International B215 and Metal Powder Industries Federation (MPIF). These sampling standards provide practical guidelines to obtain a representative sample from the whole lot, and then apply the powder characterization techniques to measure the powder properties. Further, manufacturers will be able to leverage the characterization results to pose requirements for suppliers, select the best supplier, and improve the powder re-use practices.

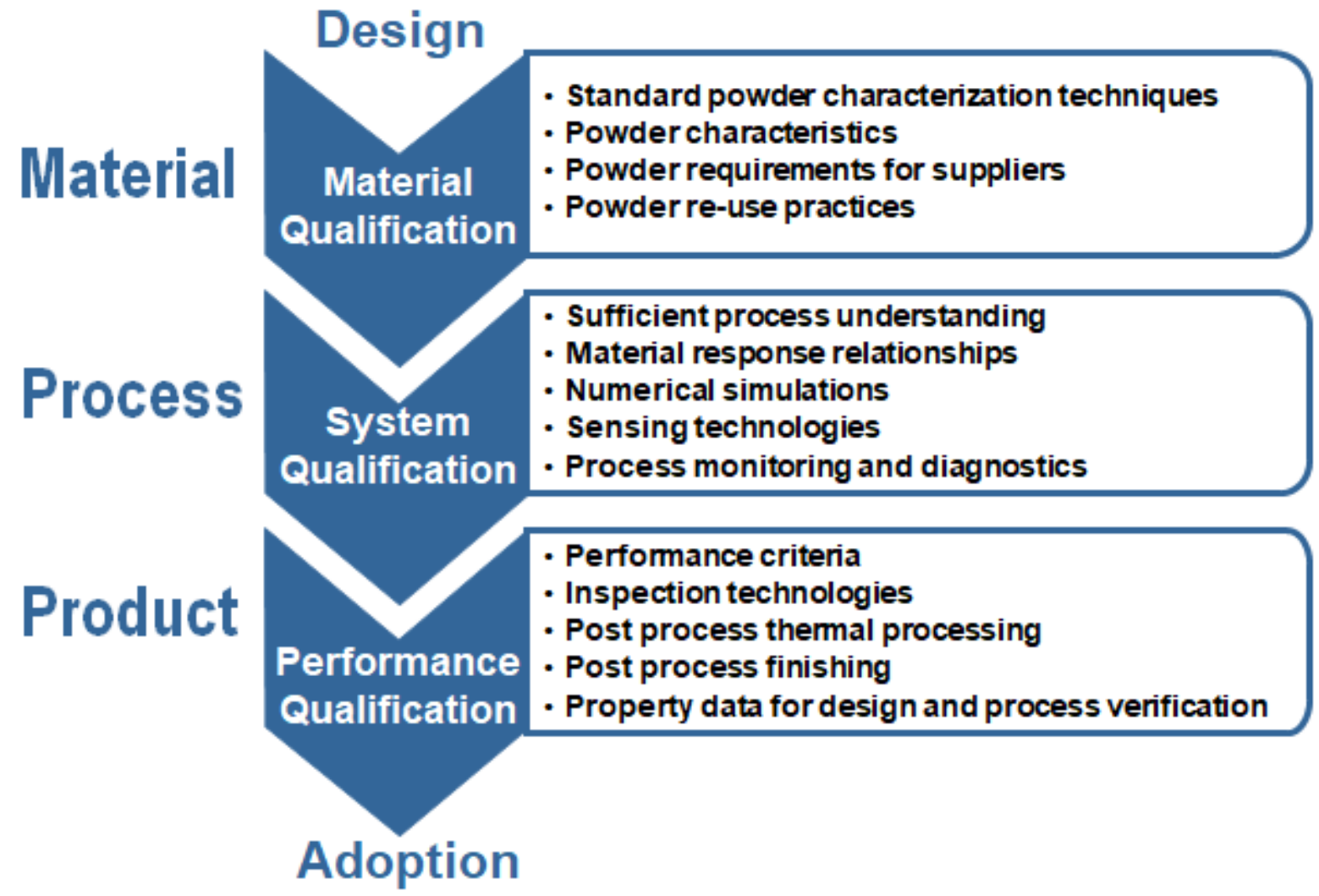

**Figure 6. AM measurements for the qualification of material, process, and product**

After pre-build material qualification, there are also system qualification in the AM process and performance qualification of the part after the build is completed (see Figure 6). In this paper, we mainly focus on the in-situ sensing of AM process performance to improve the understanding of machine-process physics, in-process monitoring, diagnostics, and prognostics. System qualification is detailed in the next section, and we briefly discuss post-build measurement and inspection in Section III.C.

## B. In-situ sensing and measurement

The in-situ sensing of AM is a rapidly developing area encompassing new hardware systems, approaches for system integration, and data analytics. The need for in-situ sensing in AM is motivated by the fact that a defect in any layer, if not detected and promptly corrected, will remain permanently sealed in on the deposition of subsequent layers. Recent review papers in this area include Grasso and Colosimo [27], Mani *et al.* [28, 29], Moylan *et al.* [30], Everton *et al.* [31] Spears and Gold [32], and Tapia and Elwany [33]. The challenges for in-situ sensing of AM are steep and discussed as follows.

(i) Each type of AM process (of which there are currently seven) imposes a unique layer bonding mechanism ranging from photochemical-initiated bonding to thermal-induced bonding; therefore, it is not possible to devise a generalized sensing scenario that is decoupled from the process physics.

(ii) The defects in AM are multifarious, and are linked to specific process phenomena that range across length scales [34]. For example, delamination and cracking in LPBF processes occur at the part-level (100 μm and above, and extending to the millimeter scale and beyond) due to thermal-induced residual stresses. In contrast, balling and keyhole melting are related to the instability at the meltpool-level (less than 100 μm). A single sensor is not likely capable of capturing these diverse phenomena.

(iii) Integrating sensors into AM machines is difficult due to the tight form factor, and mechanics of the process [35]. In the fused filament fabrication process, for instance, material in the form of a polymer filament is heated past its glass transition temperature, and deposited by a nozzle. The gap between the nozzle and the top of the part is of the order of tens of millimeters. Therefore, sensors such as an infrared thermal camera, are intractable to be mounted near the nozzle to obtain the surface distribution. This is because a large surface of the part will be blocked by the nozzle as it translates over the part [36]. A similar argument is made for the material jetting process.

(iv) In the LPBF process, layers of the powder material are spread across a bed and melted with a laser. The temperature gradient in the part is responsible for a host of defects, such as microstructural heterogeneity and delamination [37-39]. However, it is tractable only to obtain an estimate of the surface temperature distribution with the use of infrared (IR) cameras and pyrometers. The temperature at the bottom layer is not easy to obtain in LPBF because the part is surrounded by powder, which acts as an insulating medium and progressively attenuates the thermal signatures generated as the laser melts material on the layers near the top.

Moreover, it is not possible to obtain the temperature distribution in the interior of the part without altering the process flow, for example, a thermocouple can be introduced inside the part by stopping the process [40, 41]. However, this will lead to loss of the chamber atmosphere and invariably alter the thermal profile. Researchers in Penn State's CIMP-3D have pioneered wireless sensing attachments that fit into the power bed, and collect temperature information from thermocouples and strain gages [40, 41]. Moreover, the thermal phenomena in LPBF occur at multiple spatial and temporal scales. For example, the meltpool-related thermal phenomena are at the order of a few micrometers, and last for tenths of seconds, with cooling rates exceeding $10^5$ degrees per second. In the same vein, the surface-level thermal signatures last for a few seconds. Hence, different thermal imaging modalities are required for measuring meltpool-level and part-level phenomena. For the meltpool thermal imaging, a high framerate thermal camera with imaging range in the shortwave infrared region is typically used, while at the part-level, a long-wave infrared camera with a large field-of-view, smaller framerate and integration time is used [30, 42, 43].

(v) Even though the process dynamics might be notionally similar, such as DED and LPBF, the sensors from one process cannot be readily transferred between them. For example, in the DED AM process, the meltpool is several orders of magnitude larger than in LPBF, and in the former the meltpool can approach the millimeter level, while in LPBF it is close to 100 μm [44]. Likewise, the deposition rates in DED can be more than ten times that of LPBF. Moreover, in DED, the part is exposed on all sides in the chamber, thereby, convection forces (due to carrier gasses from the nozzle) and radiation are all active at the same time. Consequently, it is exceedingly difficult to demarcate and measure all of these heat transfer mechanisms.

(vi) The sensor measurements must be synchronized with the state of the process if the data is to be used for process control. Furthermore, the data from multiple sensors must be synchronized with each other. From an LPBF perspective, recording the process state would involve capturing the position of the laser (i.e., the angular displacement of the galvanometer), and merging the laser position with the sensor data being acquired. In other words, the data acquisition system must communicate with the AM machine and sensor hardware with temporal error in the microsecond range (the laser in LPBF can translate at a velocity exceeding 0.5-1.0 meter per second). The challenge is further complicated given that the sensor array may include both temporal sensors, such as photodetectors, and image-based sensors, such as thermal and optical cameras.

To overcome these barriers, researchers use heterogeneous sensing modalities [44]. A notable example of such a multi-phenomena sensing array in LPBF is the so-called Open Architecture LPBF platform at Edison Welding Institute (EWI), which is currently instrumented with the following sensors [45-47]:

(i) *Local sensors for monitoring the meltpool-level phenomena (10 μm to 200 μm scale).*

- Coaxial shortwave infrared thermal camera for meltpool temperature measurement (85 frames per second - fps, 13.4×7.12 mm field-of-view, 5 μm spatial resolution).
- Coaxial high-speed camera to track the meltpool shape (1000 fps, 10 μm resolution).
- Photodetector to record the meltpool intensity (350 – 1100 nm, 10 kHz sampling rate).
- Spectrometer to measure the optical emission in the meltpool region (200 to 1100 nm, 1 kHz).

(ii) *Global sensors for monitoring phenomena at the bulk part level (500 μm – 100 mm).*

- Coaxial short-wave infrared thermal camera focused on the powder bed to detect part temperature gradients (4 fps, 127×95 mm field of view, 400 μm resolution).
- Laser interferometer (405 nm) for measuring surface finish and distortion in a layer.
- Structured light optical imaging of the powder bed with two digital cameras to detect distortion of the part (21 fps, 25.4×14.7 mm field of view, 6.6 μm pixel resolution, 165 pixel/mm fidelity).
- Acoustic microphone and a surface acoustic wave transducer to detect when the part cracks due to distortion or makes contact with the powder re-coater (sampling rate of 10 to 40 kHz).

(iii) *sensor data acquisition, data synchronization with the laser position, and noise isolation.*

- Close to two terabytes of sensor data are acquired in a 12-hour build cycle on EWI's LPBF platform. Researchers at EWI have built the hardware and software mechanisms to ensure the seamless acquisition of sensor data of such high volume, variety, and sampling speed (a big data problem).

EWI's open-architecture LPBF platform provides the capability to measure the temperature distribution in the part and track changes of thermal gradients that are not available on other commercial LPBF systems. Another recently operational, and comparable apparatus is the *Additive Manufacturing Metrology Testbed* at the National Institute of Standards and Technology (NIST) [46]. In addition, CIMP-3D at Penn State developed a multi-sensor suite for monitoring and control of a commercial 3D System ProX 320 PBFAM system, as shown in Figure 7. The multi-sensor suite has also been demonstrated on 3D Systems ProX 200, EOS M280 and GE Concept Laser M2 machines. The system consists of a variety of sensors as follows:

- High-resolution/high-magnification imaging system (6 differing lighting schemes)
- Two (2) high-speed/high-magnification cameras, including a coaxial camera with 405 nm filter and a front-facing camera with 520 nm filter
- High Speed Video (>33,000 fps)
- Optical Process Emissions (100 kHz), including spectrometer and multi-spectral sensors
- Acoustic sensors (100 kHz)
- Thermal imaging and DMP Melt Pool sensor

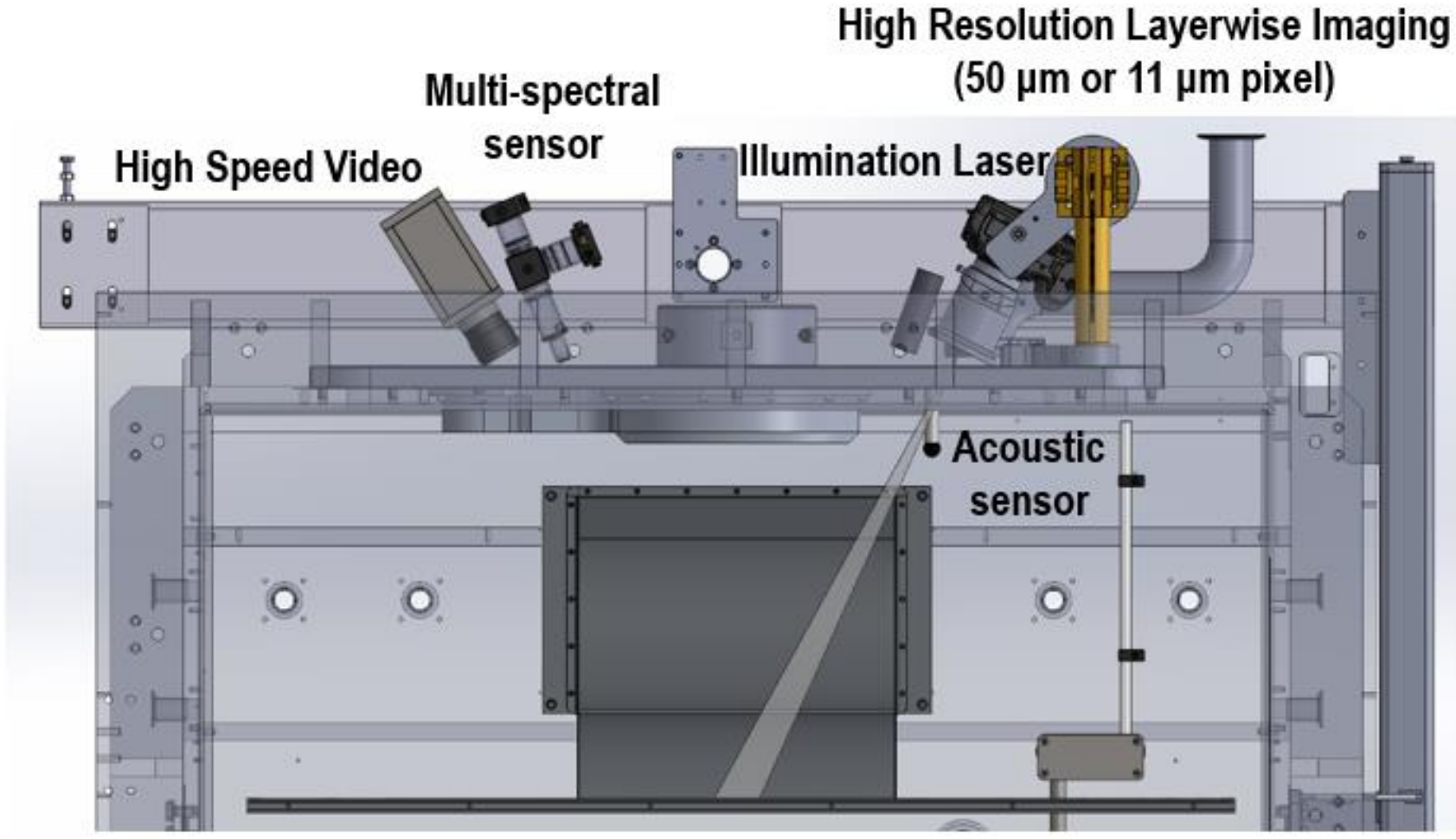

**Figure 7. An illustration of multi-sensor Suite for monitoring a Commercial ProX 320 PBFAM system**

This multi-sensor suite includes an optical layer-wise imaging system to monitor the LPBF AM process, which consists of a 36.3 megapixel DSLR camera that is placed inside the chamber of EOS M280 machine [12]. In-process optical images have also been collected and used to identify and characterize defects caused by lack-of-fusion in the LPBF process [48]. Nassar et al. describe the use of an in-situ optical emission spectroscopy system consisting of two filtered photodetectors in a series of papers [10, 49, 50].

A recent paper demonstrated the use of this relatively inexpensive system to monitor lack-of-fusion porosity in Inconel 718 test parts [1]. Inconel has Chromium as an alloying element. When Inconel is fused (melted) by the laser, atomically-excited chromium is vaporized and emits photons corresponding to electronic transitions. One set of transitions occur wavelength around 520 nm [51]. The rationale is that if melting is stable, line emission from the vapor will also be stable. A key innovation is the use of two photodetectors, one of which is filtered to have a frequency spectrum in a region where line emissions are not likely, and measures emissions pertaining to the background radiation (a wavelength different from the line emission wavelength, called the continuum emission spectrum). Further, Nassar *et al.* divide this difference (line emission intensity minus continuum emission intensity) by the continuum emission intensity; this ratio is called the line-to-continuum ratio. Nassar *et al.* use features derived from the line-to-continuum ratio as inputs to detect lack-of-fusion porosity [1]. Notably, multi-sensor systems generate high-dimensional and heterogeneous data (e.g., time series, video, image profiles) that provide rich information about AM processes. However, realizing the full potential of these data for AM system qualification depends to a great extent on the development of analytical methods to characterize, represent, and extract useful information about the defective state in each layer of AM builds, detailed in Section IV.

## C. Post-build measurement and inspection

As shown in Figure 8, post-build quality inspection and function integrity assessment for AM are often performed with radiographic-based Computed Tomography (CT). Here, CT scans of AM builds are collected with a GE vTomex M300 microfocus X-ray CT scanner, and are processed using the Volume Graphic myVGL3.0 software to extract the 2D image profiles of every layer in an AM build. The resolution of image profiles is determined by CT voxel size, typically with a pixel size of 10-50 µm or less. These data will enable the investigation of the effect of design parameters or LPBF process settings, e.g., hatch spacing (H), scan velocity (V), and laser power (P), on the defect patterns in AM image profiles. The sensor data and offline CT scans can be used to create a library of (sensor) patterns that correlates to specific defects using sensor fusion and predictive analytics. These sensor signal patterns, which exemplify specific process defects, can be integrated with prescriptive models (i.e., for decision-making) to optimize the selection of corrective action in case of an anomaly is detected in the process. The focus is to minimize defects, delamination and warpage of the final workpiece, and maximize final strength and fatigue resistance. In addition, other equipments such as coordinate measuring machines (CMM) and surface probing machines provide important information about part dimensional metrology and surface roughness [52, 53].

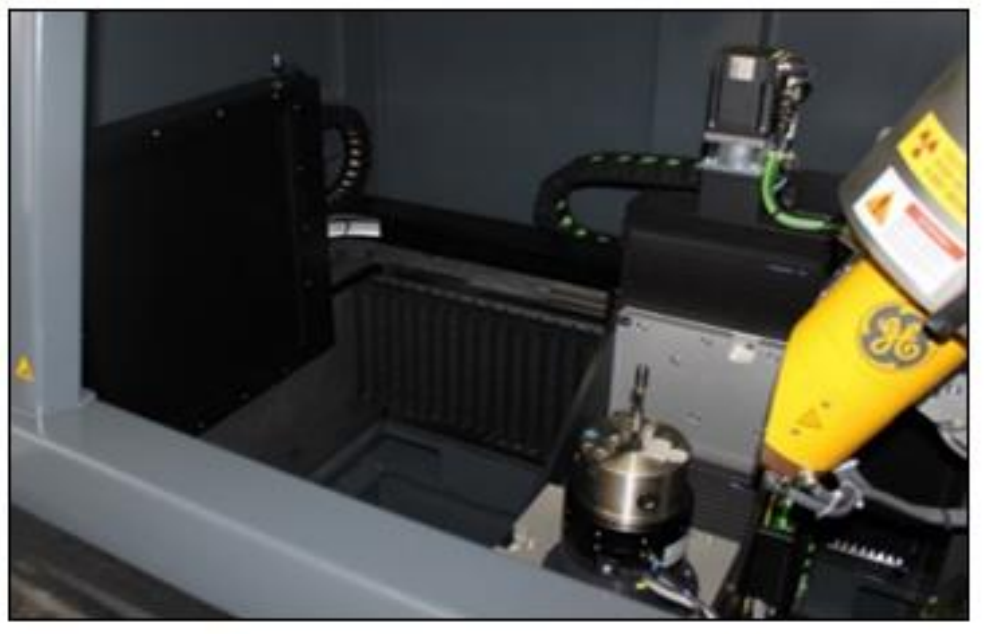

**Figure 8. Radiographic-based computed tomography for post-build inspection**

## D. AM data management

Large amounts of data are generated, exchanged, and used dynamically during AM test coupon and part development processes. As the volume of data grows with increased in-situ sensing and nondestructive examination (NDE), the types of data generated by AM activities also become richer. The information necessary for AM process qualification includes not only measurement data, but also material/machine specifications, design models, control, and management data. Characterizing the entire AM process demands a comprehensive analysis of all the information collected through the build history of thousands of parts and coupons, in the context of the complete AM value chain. As a result, it requires an effective and efficient AM data management system to ensure that data are captured, stored and used appropriately.

In the area of data management, several AM information management systems are available as commercial products. For example, the Senvol Database (http://senvol.com/database/) provides researchers and manufacturers with open access to industrial AM machines and materials. Granta, a material information management technology provider, offers the product GRANTA MI: Additive Manufacturing, specifically customized for AM data capturing and use. At the same time, multiple database and data management systems are built to organize and manage the data generated from research and industry projects. The DMSAM (Data Management System for Additive Manufacturing) was developed by researchers at Penn State's CIMP-3D (http://www.cimp-3d.org/datamanagement). DMSAM is a schema-based software tool that stores and tracks all of the data and information related to an AM part, including the state of associated AM resources (e.g., powder, software, machine), part requirements for sponsors, 3D solid models, part workflow, build plan, post-processing plan, and all data associated with part properties, in situ monitoring, post-processing, testing, and inspection. DMSAM stores data locally, communicating with global (i.e., shared) databases and generating build reports for QA/QC as needed through XML as well as Excel. NIST's Additive Manufacturing Materials Database (AMMD) [54], is a data management system built with NoSQL (Not Only Structured Query Language) database technology and provides a Representational State Transfer (REST) interface for application integration. The database captures rich research data sets generated by the NIST AM program (https://ammd.nist.gov/) based on an open XML schema. In addition, as an open data management platform, the AMMD system is set to evolve through co-development of the AM schema and contributions of data from the AM community.

However, due to the multifarious factors that could affect AM part quality, existing data-driven AM process qualification requires extensive testing of material which is beyond the capability of any individual organization. None of the existing databases provide comprehensive data sets with a multitude of geometries and processes settings by itself to qualify an AM process for parts with various features and specifications. In order to significantly reduce the cost and time associated with the data generation for AM process qualification, a collaborative data space is required and a collaborative data management system is necessary. Figure 9 shows a multi-tier AM collaborative data management system that is featured by: a) distributed data storage facilitated by using common data terms and definitions, b) collaborative linked data through federation based on neutral data formats, c) continuous knowledge management by extracting AM material process-structure-property relationship automatically from AM data, d) lifecycle and value chain based decision support, and e) an adaptive data generation system that helps AM community to efficiently design experiments. The collaborative data management system is set to identify, generate, curate, and analyze AM data through AM product lifecycle and can significantly reduce the cost and time associated with AM product deployment.

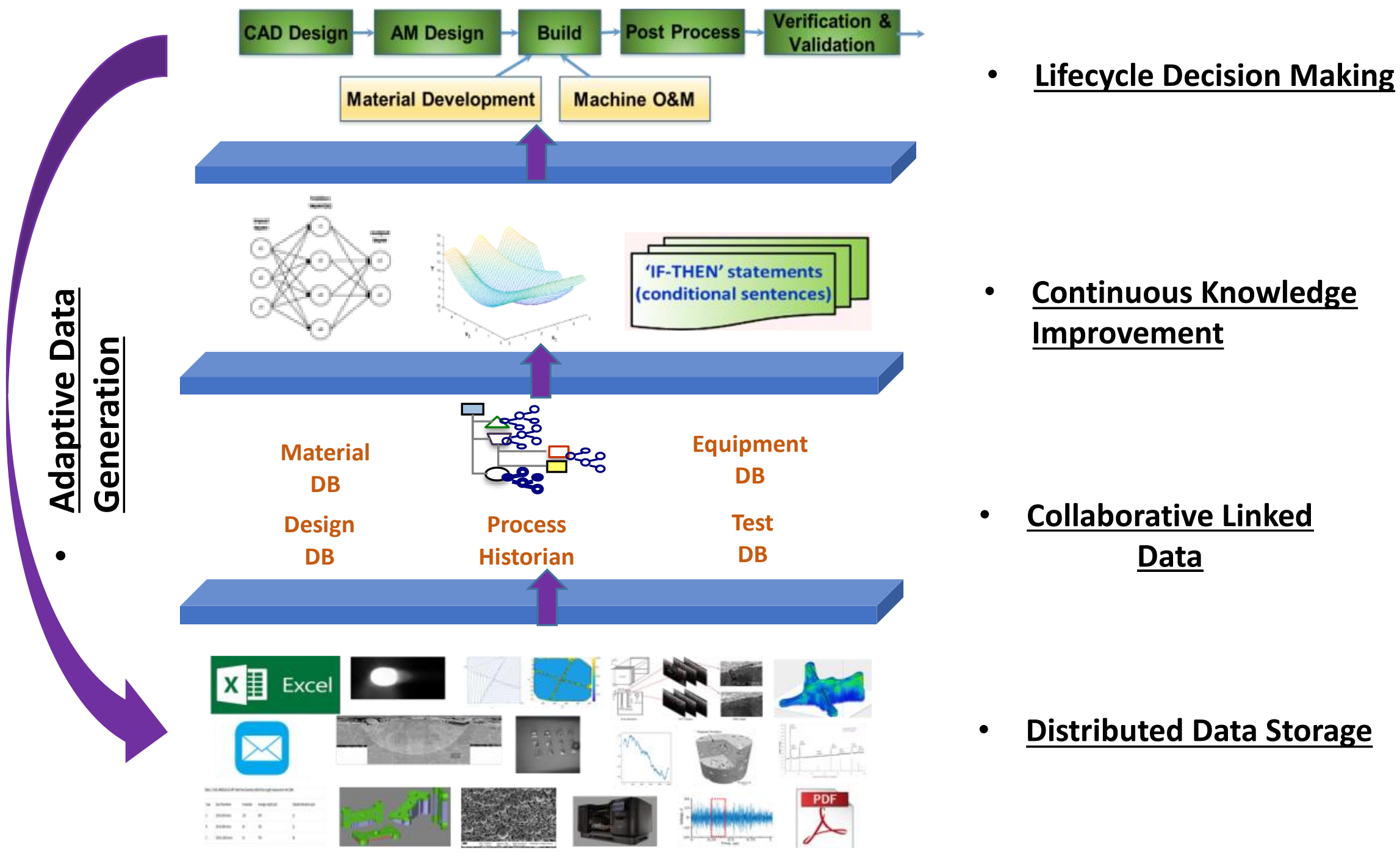

**Figure 9. Multi-tier Data Management System for Additive Manufacturing**

## IV. Analyze the Data

The "analyze" step focuses on the extraction of useful information from online and/or offline data collected in the "measure" step, or from historical data available in the AM data management system. The main purpose is to explore the interrelations among key variables (i.e., process inputs, outputs, and in-process variables) during the AM process, model causal relationships between these variables and quality problems, and further develop new understanding about how they contribute to process variability and product defects. In other words, multiple sources of variability may exist in the AM process and can potentially lead to quality problems in products and customer services. The "analyze" step helps delineate and determine the random causes and assignable causes to the quality problems. If only random causes (i.e., non-assignable factors, not identifiable) are presented in the process, then the distribution should be normal. However, if there are assignable causes, then "analyze" tools should be able to monitor the process, and detect when and how the process performance is affected. As such, the process can be stopped to look for assignable causes and eliminate them to resume the normal production.

However, advanced sensing systems bring more and more complex-structured data from the "measure" step for AM quality management, which are different from geometric features, linear and nonlinear profiles generated in conventional manufacturing settings [55]. For example, CT scanning and layerwise imaging result in high-dimensional image profiles from the AM process. As such, traditional "analyze" tools such as control charts and confidence intervals are limited in the ability to handle such high-dimensional image profiles. Control charts and confidence intervals are much easier to establish for a single random variable or multiple variables (e.g., geometric features of products) in the setting of mass manufacturing, but are more difficult to be developed for high-dimensional images, let alone geometrical structures in these images may vary from one layer to another layer in the AM builds. Hence, new "analyze" tools are urgently needed to help handle and connect large amounts of data, model the cause-and-effect relationships among key process variables, and pinpoint potential root causes to quality problems during the AM process. This, in turn, will help the "improve" step (see Section V) to further identify and develop new strategies for quality improvements. New experiments can then be designed to test the effectiveness of these improvement strategies on either physical AM machines or computer simulation models

## A. Engineering design vs. build quality

Engineering design and relevant parameters are one of key process input variables during the AM process. Traditional subtractive manufacturing tends to be limited in the ability to handle complex designs. "Design for manufacturing" refers to the conventional scheme that adapts a design to enable manufacturing within the capability of available machine and tools. AM offers a higher level of design freedom and enables the new scheme of "manufacturing for design". Complex designs can now be manufactured in a layer-upon-layer fashion with the new generation of AM technology. Nonetheless, complex designs still pose quality challenges on AM-fabricated products, despite the fact that AM can handle certain aspects of fabrication better than traditional manufacturing technologies.

The new research question is whether and how design parameters influence the quality of AM builds? Our prior work has designed and performed experiments on an LPBF machine to investigate how design parameters (i.e., height, width, recoating orientation, and hatching pattern) impact the quality in the final build of thin-wall structures [56, 57], which are widely used in the fabrication of heat exchangers. As shown in Figure 10, our experiments built the thin-wall structures with a variety of design parameters, i.e., heights, widths, recoating orientations, and hatching patterns. The metal power is Spherical ASTM B348 Grade 23 Ti-6Al-4V, with the distribution of powder size in the range $14\mu m - 45\mu m$. Each build includes 25 thin walls that are fabricated on a $15mm \times 15mm \times 55mm$ platform. Experimental factors such as height, width, recoating orientation, and hatching pattern are detailed as follows:

- **Height:** The height of thin walls varies from 0.6mm to 3.0mm with the step size of 0.1mm. The height-to-width ratio is 0.1 in each thin wall.
- **Width:** The width of thin walls ranges from 0.06mm to 0.3mm with a step size of 0.01mm.
- **Orientation:** Thin wall structures are fabricated vertically upwards with the layer thickness of $60\mu m$ in 3 orientations with respect to the travel direction of recoater blade (i.e., $0^o$, $60^o$, and $90^o$). Figure 10 shows 3 orientations with the reference of recoating direction.
- **Hatching:** The hatching patterns of thin walls follow the standard processing path of EOS machines, but various categories of scan paths are utilized when the width increases (see Figure 10). Fin 1 and 2 have two inner rectangle paths, two outer layer paths (or contours), and rotating diagonal hatching from rectangles. Fin 3-14: there are three outer layer paths and rotating diagonal hatching inside the innermost rectangle. Fin 15-18: there is one rectangular hatching. Fin 19-25: there is one thin area path.

As shown in Figure 10, we fabricate three thin-wall parts in this experimental study, each of which includes 25 thin walls. The orientations are different for three thin-wall parts on the build plate. In other words, the orientation of each thin-wall part is adjusted to the degree of $0^o$, $60^o$, or $90^o$ with respect to the travel direction of recoater blade in the EOS machine. After fabrication, we scan each build with X-ray computed tomography (XCT). These XCT images are then registered with the original CAD models to extract the quality characteristics (e.g., edge roughness, defect levels) in each layer of the thin wall. Here, the edge roughness refers to the geometric deviation of build boundary between CT scans and CAD designs. The defect level refers to the number and degree of defects in each layer of the thin wall. These quality characteristics are tracked from one layer to another for the detection of impending collapse of thin-wall failures. See [56, 57] for the analysis of variance with respect to design parameters.

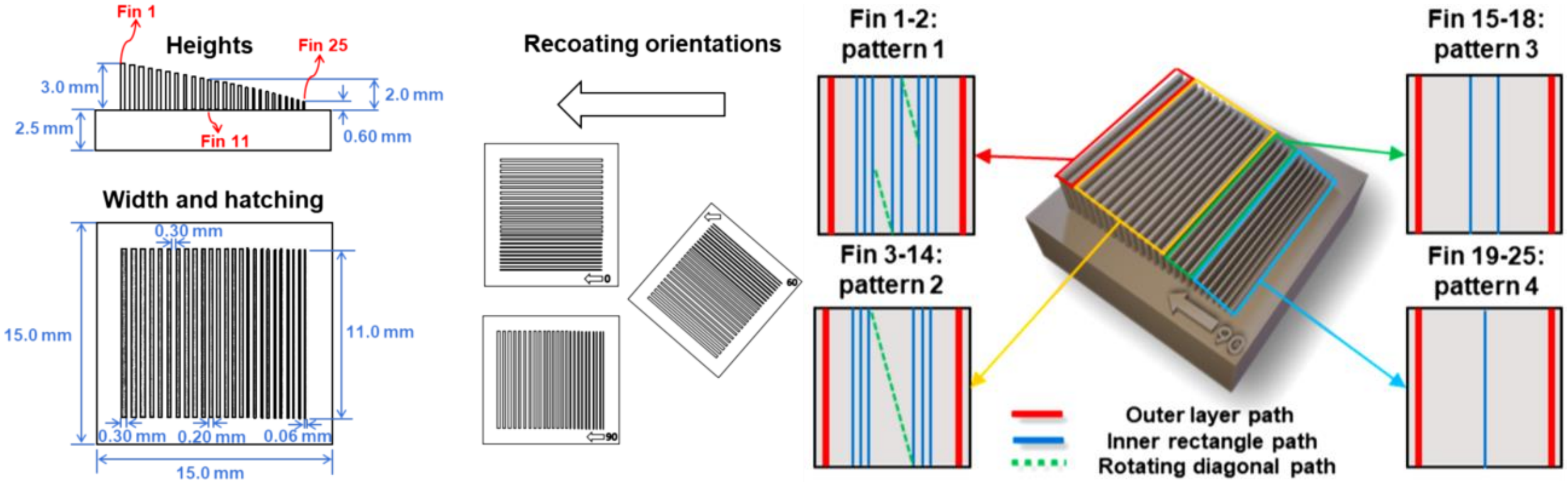

**Figure 10. An illustration of design parameters (i.e., orientation, width, height and hatching pattern) for the thin wall structure**

Through the analysis of XCT data and in-process imaging data, experimental results show that the build quality of thin-wall parts is impacted by design parameters (height, width, height-to-length ratio) and machine settings (hatching and recoating orientation). This study helps provide a set of design guidelines on the use of LPBF machines for the fabrication of thin-wall structures as follows.

- The $0^o$ orientation gives a superior quality in the thin-wall builds than other orientations. Fewer defects are generated when the travel direction of recoater blade is parallel to the long edge of a thin-wall. It should be avoided to build thin-wall structures in the $90^o$ orientation, which tends to generate more flaws by making the recoater motion perpendicular to the long edge of a thin-wall.
- The height of a thin wall should not be more than nine times its width. Otherwise, this thin-wall build tends to collapse. The LPBF machine in this experiment is limited to build the thin-wall structures with the width that is smaller than 0.15mm. If the length-to-width ratio exceeds 73 (11 mm / 0.15 mm), thin walls also tend to collapse.

This study made an attempt to answer the research question about whether and how design complexity influences quality characteristics of AM thin-wall builds? There is more research to be done to optimize the engineering design for AM. For example, how to generalize the design guidelines for different LPBF machines, process conditions or thin walls with overhang structures?

## B. Machine setting vs. build quality

Machine settings (e.g., hatching space, laser power, scan velocity) often influence final outcomes of AM manufacturing process, including the cosmetic appearance and build quality. To increase the information visibility and cope with the complexity in machine-process interactions, advanced sensing is increasingly employed in AM (see the multi-sensor suite and CT scanner in Figure 7 and Figure 8), thereby generating large amounts of data (e.g., optical images and post-build CT scans). Realizing the full potential of sensor data hinges on the development of new statistical quality control (SQC) methods. Existing SQC methods for conventional manufacturing processes are more concerned about key features of finished products (e.g., dimensional accuracy), linear and nonlinear profiles, as opposed to high-dimensional sensor data. The research on AM sensing, machine-process interaction, and QA/QC poses several new challenges:

- **Sensor-based metrology for in-situ quality inspection:** Traditional QA/QC techniques such as surface metrology geometric and dimensioning and tolerancing (GD&T) are more concerned about Euclidean features of the finished AM products. They are offline and not amenable for the inspection of internal defects in AM parts with complex geometries [58-61]. In the absence of sensor-based approaches for in-situ quality monitoring, benchmarking of AM builds remains relegated to post-build inspection and qualitative attributes [62-64].

- **Statistical quality management for AM:** Current quality monitoring approaches are (i) offline; (ii) based on purely data-driven techniques (neural networks, mixture Gaussian modeling, statistical analysis); or (iii) lumped-mass formulations [65-68]. Very little has been done to investigate AM quality management using sensor-based analytical models and layerwise AM QA/QC strategies. In-situ monitoring provides an opportunity to in-process AM defect mitigation that is indispensable for manufacturing industries mandating stringent quality standards and product aesthetics.

Hence, the first step is to extract useful information from AM sensing data and then estimate the defect levels in AM builds. As shown in Figure 11, fine-grained images of AM builds demonstrate fractal characteristics over a range of scales. Traditional linear methods are limited in the ability to handle nonlinear fractals and irregular patterns in the images. Fractal analysis extracts a single fractal dimension that describes the self-similarity (scale-invariant) behavior of fractal objects, but cannot fully characterize multifractal patterns that are often shown in real-world objects [69, 70]. However, image profiles of AM builds often comprised of complex self-similar patterns that are not due to a single fractal but rather the existence of a spectrum of fractal dimensions. These fractals interact with each other and then generate highly nonlinear and complex self-similar behaviors (see Figure 11).

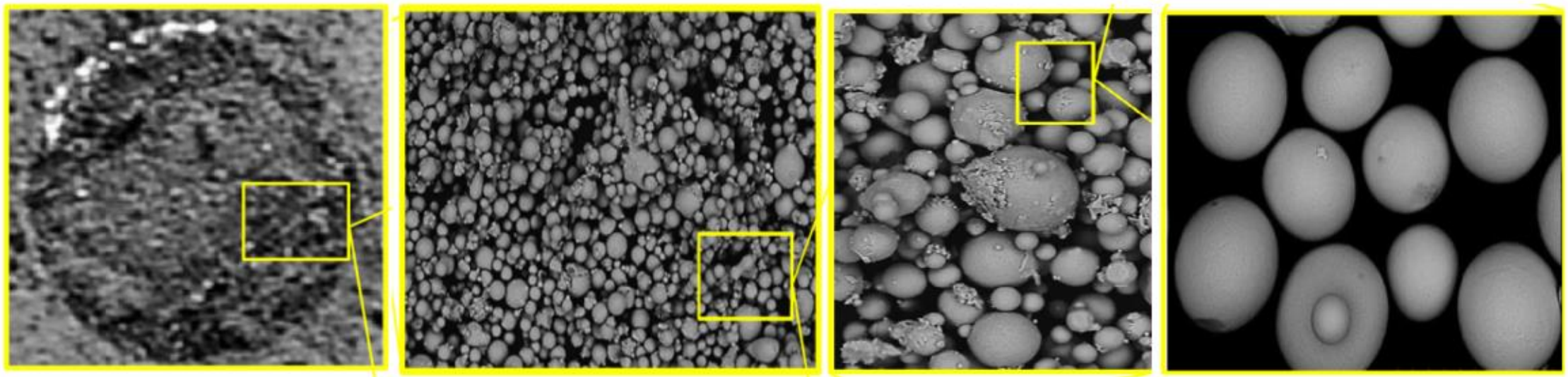

**Figure 11. An illustration of multifractal patterns in the image profiles of an AM build**

Little has been done to characterize multifractal patterns in large amounts of image profiles to investigate how machine settings influence the AM build quality. Our prior work has developed new multifractal methods for the analysis of large amounts of AM imaging data and extracts features that are sensitive to the defects, instead of extraneous factors and random noises [69-73]. As shown in Figure 12, multifractal analysis characterizes the nonlinear and self-similar behaviors of AM images in multi-scale lenses, ranging from large-scale approximations to small-scale details. AM images are then decomposed as an interwoven set of fractals with different dimensions, which is shown as the multifractal spectrum. In addition, lacunarity measures the degree or extent on how this set of fractals fill the space, which cannot be provided by multifractal analysis alone. As such, we developed the method of joint multifractal and lacunarity analysis to characterize and quantify the nonlinear and multifractal patterns in AM images that cannot be otherwise achieved by either traditional statistical methods or fractal analysis.

After the multifractal characterization results of AM images, we investigated how AM machine settings (e.g., laser power ($P$), hatch spacing ($H$), and velocity ($V$)) influence the build quality. In the experimental study, we printed cylinder parts in the EOS M280 machine with varying levels of machine settings (also see Figure 20). Specifically, laser scanning velocity is increased from 1250 mm/s, 1562.5 mm/s, to 1875 mm/s. Hatching space is varied from 0.12 mm, 0.15 mm to 0.18 mm, and laser power is decreased from 340 W, 250 W, to 170 W. Further, a regression model is constructed to predict the relationship between machine settings with the Hotelling $T^2$ indices of build quality, which are computed with multifractal and lacunarity features of XCT image profiles [69-73]. The model achieves the adjusted R-squared value of 94.76%, showing strong correlation between process conditions and build quality.

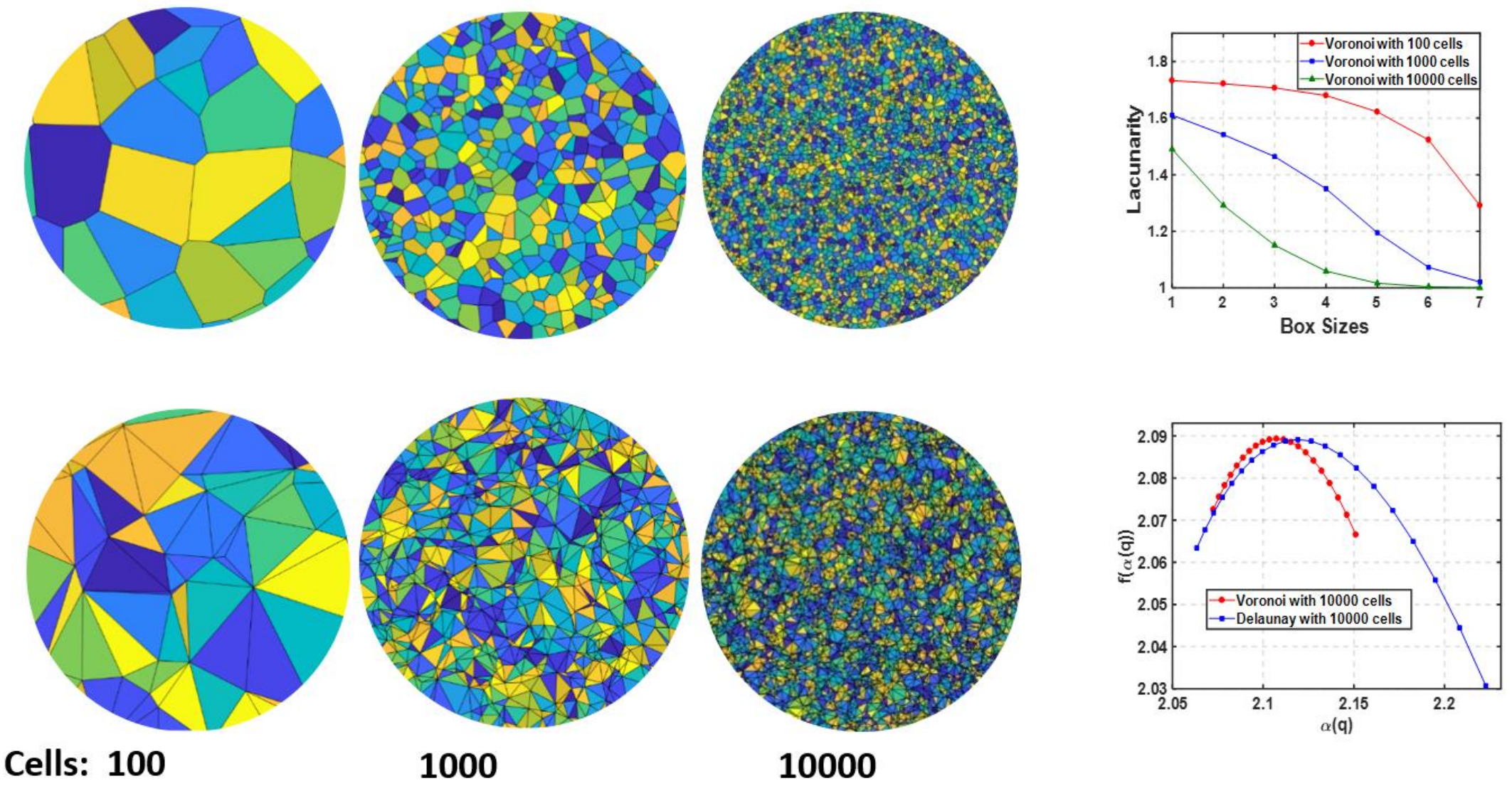

**Figure 12. Multi-scale analysis of fractal and lacunarity patterns in the layerwise AM images with Voronoi tessellation from 100 cells, 1000 cells to 10000 cells; and Delaunay triangulation from 100 cells, 1000 cells to 10000 cells, for multi-resolution quality inspection of the layerwise AM build.**

## C. In-situ sensing variables vs. build quality

CT scans help characterize the quality of a finished build, but cannot detect the flaws during the AM process. In-situ sensing provides a means for on-the-fly defect characterization. As shown in Figure 13, a drag link part with complex geometry was printed in CIMP-3D with intentional defects at four layers (i.e., 1.5mm, 6.7mm, 12.0mm, and 16.0mm), each of which includes 8 defects as follows:

- 0.050 mm, .250 mm, .500 mm, and .750 mm cubed defects are on each plane.
- 0.050 mm, .250 mm, .500 mm, and .750 mm diameter cylinder defects are also on each plane surrounding the cubes. All cylinders have 1:1 diameter to depth ratio except for the .050, which has a depth of .250 mm.
- The top of the defect is the flat plane in the build direction.

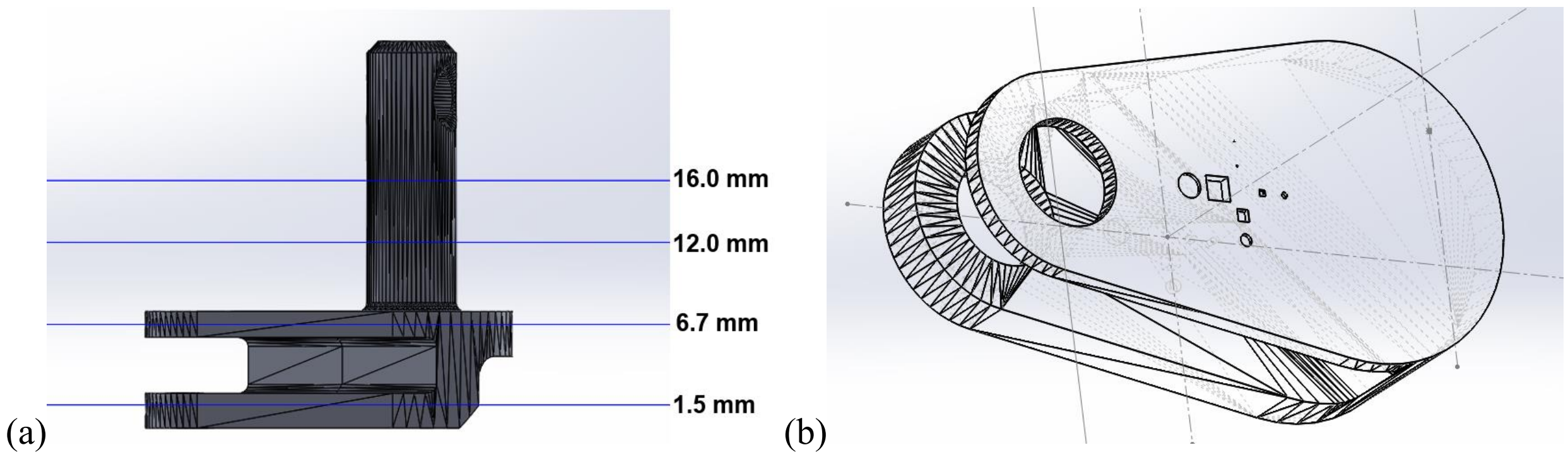

**Figure 13. (a) Four layers of intentional defects; and (b) different shapes and sizes of defects**

In-situ optical images are recorded after each layer is printed. This experimental study is aimed at predicting incipient defects from in-situ imaging data for quality control in AM processes. In the state of the art, deep neural networks (DNN) models show superior performance in the handling of imaging data. However, layerwise imaging data from AM processes pose significant challenges on DNN defect analysis:

- **Region of Interests (ROI):** Each image of the build plate consists of metal powders and many parts. As such, there is a need to delineate the image for a specific part. Often, a squared region is cropped around the part and then images of layers are fed to the DNN model. This guarantees the same dimensionality of input images to the DNN model. However, due to the broad geometrical diversity from one layer to another, images of some layers will have small part geometries and large powder areas while others have large part geometries and small powder areas. DNN learning will be biased by the layerwise geometrical diversity, as well as the varying areas of unfused powders. Therefore, it is more desirable to leverage CAD files to delineate and register the ROI for the part geometry in each layer (also see Figure 14).
- **Layer-to-layer geometry variation:** AM provides a higher level of flexibility for the low-volume and high-mix production, even once for a one-of-a-kind design. As shown in Figure 5, AM fabricates the build directly from a complex CAD design through layer-upon-layer deposition of materials. Although we may register the ROI for the part geometry in each layer, there will be ROI variations among layers. Hence, both the shape and dimensionality of ROIs will be varying from one layer to another. The inconsistent ROIs consist of different numbers of pixels, and cannot be used as inputs to the DNN models for learning the incipient defects in the layers.
- **ROI segmentation and spatial characterization:** To tackle the challenge of inconsistent ROIs, one approach is to extract features from layerwise ROIs (e.g., mean, median, variance). However, statistical features tend to aggregate useful information within the ROIs, thereby leading to the deficiency in defect characterization and predictive modeling. The other approach is to segment ROIs into smaller regions of interest (sROIs) with the same number of pixels. Although the dimensionality of ROIs is changing from one layer to another, the greatest common divisor (GCD) for ROIs of all layers can be leveraged to segment sROIs with the same number of pixels. However, these sROIs may still have variations in shapes. Further, spatial characterization can be used to measure spatial correlations among pixels and describe pertinent patterns about defects in the sROIs. As the number of pixels is consistent in the sROIs, characterization images share the same dimensionality that can be fed into DNN models for the learning and prediction of defects in each sROI in the AM processes.

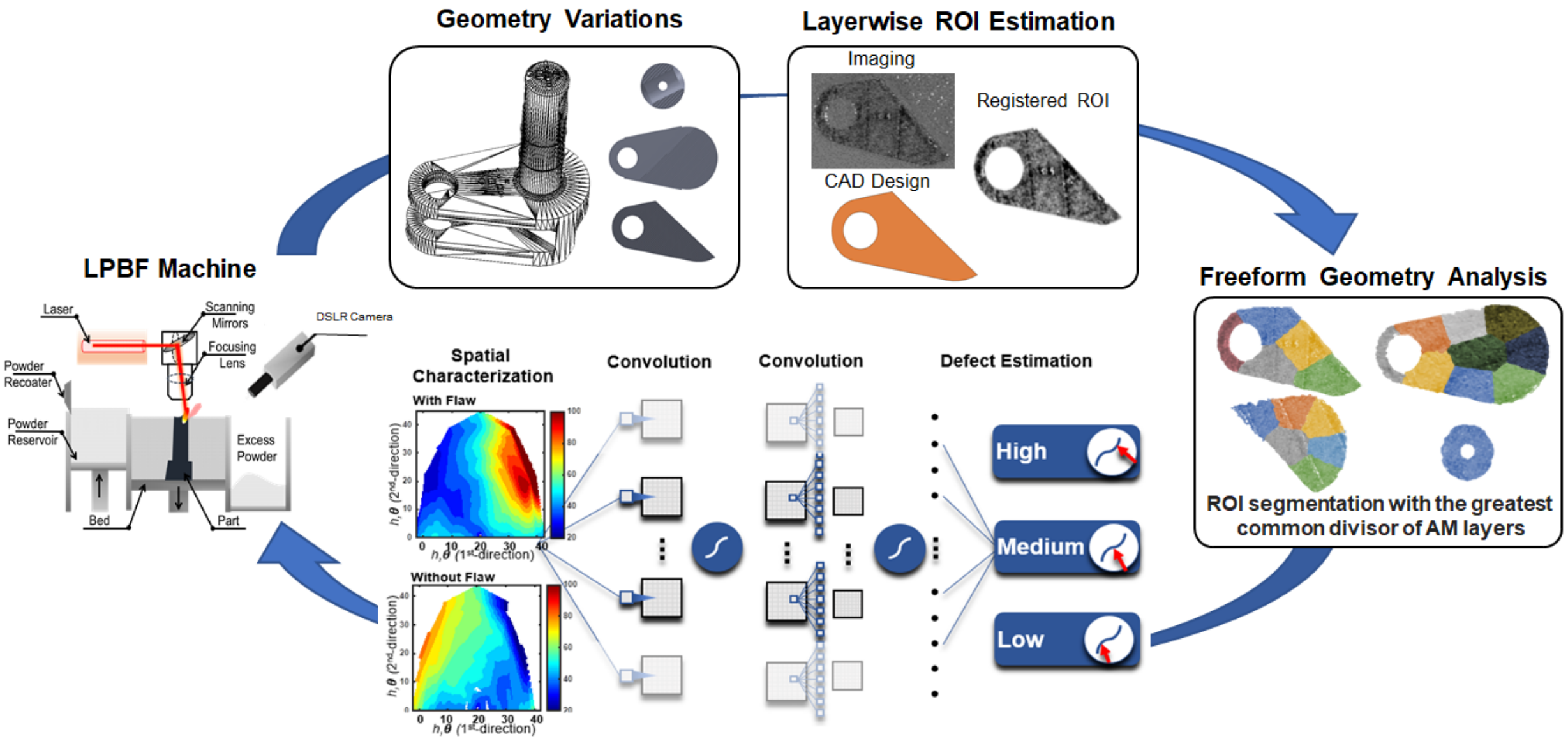

**Figure 14. A schematic illustration of Deep learning of incipient defects from in-situ image profiles.**

As shown in Figure 14, our prior work has designed a new DNN model to learn incipient defects from sROIs of in-situ image profiles [74, 75]. The experimental study provides large amounts of images taken for each layer with different lighting schemes. To tackle the aforementioned challenges, DNN learning of in-situ AM defects consists of the following critical steps:

***(1) Image registration and segmentation***: We first used the CAD design to perform shape-to-image registration and extract the ROIs of 362 layers in the drag link part. Then, these ROIs are segmented into 1,708 sROIs, each of which has the same number of pixels. Further, the dyadic partitioning of sROIs can be used to split each region into smaller subregions and provides a large amount of data for multi-resolution DNN learning of layerwise AM defects.

***(2) Spatial characterization:*** Although these sROIs are in different shapes, we utilized the spatial characterization to extract pertinent patterns about defects from sROIs and then fed images of spatial correlations for the deep learning.

***(3) Deep learning:*** The DNN model includes a series of convolutional layers to learn sROI characterization images with multiple levels of abstraction. Each hidden layer is followed by non-linear modules, which transform the representation at one level into a representation at a higher, slightly more abstract level [74]. The DNN builds up effective learning and representations of various intentional defects (i.e., embedded in the drag link part, see Figure 13) that help significantly reduce the size of state space and state-action pairs for predictive modeling and optimization in the following subsections.

The proposed DNN model is shown to effectively predict the layerwise defects with the specificity 93.85±0.83%, sensitivity 90.01±1.56%, negative predictive value 93.83±0.67%, positive predictive value 90.03±2.34% and accuracy of 92.50±1.03%. This experimental study avoids the use of DNN as a blackbox by just feeding cropped images of layers (i.e., with the broad geometrical diversity) into the neural networks and then letting artificial intelligence (AI) figure out the defects. Instead, engineering domain knowledge is indispensable to pre-processing AM training data and developing effective AI methods for in-situ AM defect learning and analysis.

## V. Improve the System

This section presents a set of statistical methodologies – ontology models, design of experiments and simulation analysis - for the quality improvement of AM processes. The “measure” step provides rich data about key variables to increase the information visibility during the AM process. The “analyze” step extracts useful information from the data and performs the cause-and-effect analysis between and among these key process variables. Now, the “improve” step exploits data-driven knowledge to look for changes or parameter designs that can be made to the AM process so that the performance can be improved.

Ontology provides a high-level map that is useful to explore and understand the interrelationships of parameters, elements, and variables during the AM process. Hundreds of terms may be involved in the AM process ontology to describe input-output parameters of laser, thermal, microstructure, and mechanical properties of AM parts. These terms may be physical parameters or concepts that are based on mathematical modeling and physical phenomenon characterizing the AM system. For instance, the laser source affects the thermal behavior and microstructure evolution during an AM process [76], and the thermal distribution of the heat source affects the microstructure behavior and mechanical properties [77]. As a result, ontology models relate process parameters to mechanical properties and material characteristics, and can be used for process redesign, sensor selection, and quality improvement.

Furthermore, design of experiments is one of the most widely tools for quality improvement. Note that the “Analyze” step delineates multiple sources of variability in the AM process, e.g., assignable causes or random causes. Therefore, the “improve” step can then choose experimental factors and vary the factor levels with statistical designs (e.g., randomized block design, factorial design, response surface design) to investigate how these factors influence the quality of AM process and final builds. Most importantly, optimal factor settings can be determined to ensure the desired performance of AM process can be achieved, which is robust to uncontrollable factors and/or random noises [22].

It should be noted that designed experiments can be conducted on physical AM machines, or computer simulation models, or both to improve the performance of AM processes. Simulation analysis involves the design of computer experiments that is often faster and cheaper than physical experiments. As such, before expensive experiments are undertaken on AM machines, simulation analysis can help screen the process variables to reduce the number of factors and design more cost-effective experiments in the "improve" step. If the AM process is far from the desired level of performance and keeps producing a large number of defective builds, then it may be necessary to abandon the old process and redesign a new AM process. In this way, the "improve" step is converted to be a "design" step in the DMAIC approach.

## A. Ontology modeling

As shown in Figure 15, the growing body of AM research exists in many forms (e.g., papers, models, simulations, graphs, data) and is both specific to a given AM process and generalizable to AM more broadly. Several complementary efforts are underway to develop data management systems by NIST*, CIMP-3D†, Granta‡, and many others. Also, numerous sensing capabilities (e.g., photodiodes, cameras, pyrometers, thermocouples, spectrometers) are available for metal AM processes (e.g., powder bed fusion, directed energy deposition). Different sensors have been installed on different AM systems to generate empirical data to help validate simulation models, for instance, or develop process maps for different AM systems and materials. The challenge now lies in the integration of all the information into useful AM knowledge, which includes the selection of right sensors to generate the right data for the right analytics for QA/QC.

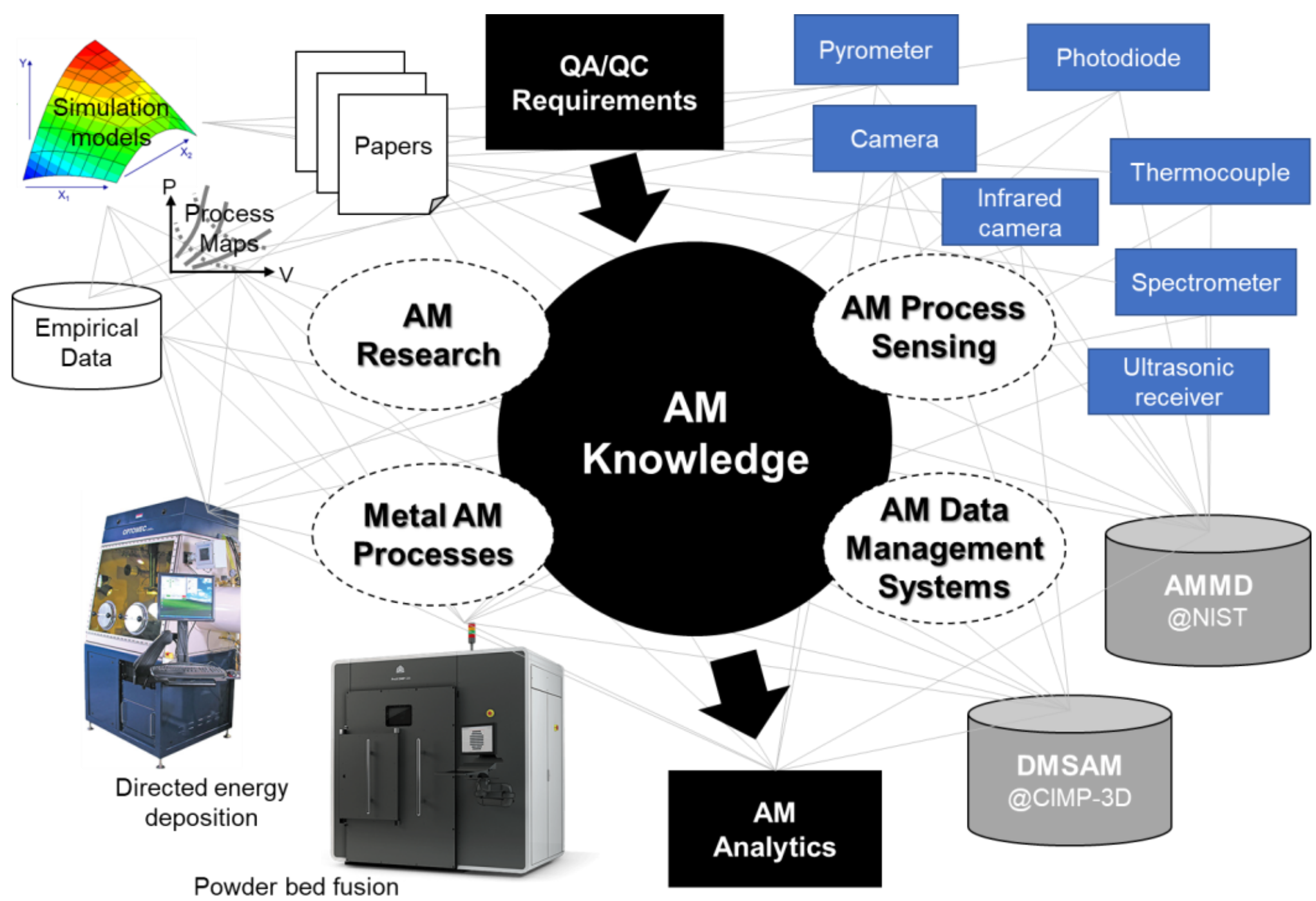

**Figure 15. Challenges navigating research, sensing, and data management for metal AM**

Our previous work developed an ontology to support AM process model development and reuse [78, 79]. The AM ontologies sought to overcome pertinent challenges about disparate AM process models and simulations (e.g., with variations in the input/output specification), not to mention levels of detail, fidelity, and composability. This limitation restricts their reuse and makes it difficult to integrate different models from different groups into the most accurate AM simulation model or for different use cases. The AM ontologies developed by Penn State and NIST allowed users to navigate complex relationships and understand the connections between different process parameters, microstructural characteristics, and mechanical properties for AM parts. A sample of the ontology is shown in Figure 15 where details on the class hierarchy for AM thermal models can be seen along with the definition of the Absorbed_laser_power class.

* https://ammd.nist.gov/
† http://www.cimp3d.org/datamanagement
‡ https://grantadesign.com/industry/products/granta-mi/support-materials-engineering/granta-miadditive-manufacturing/

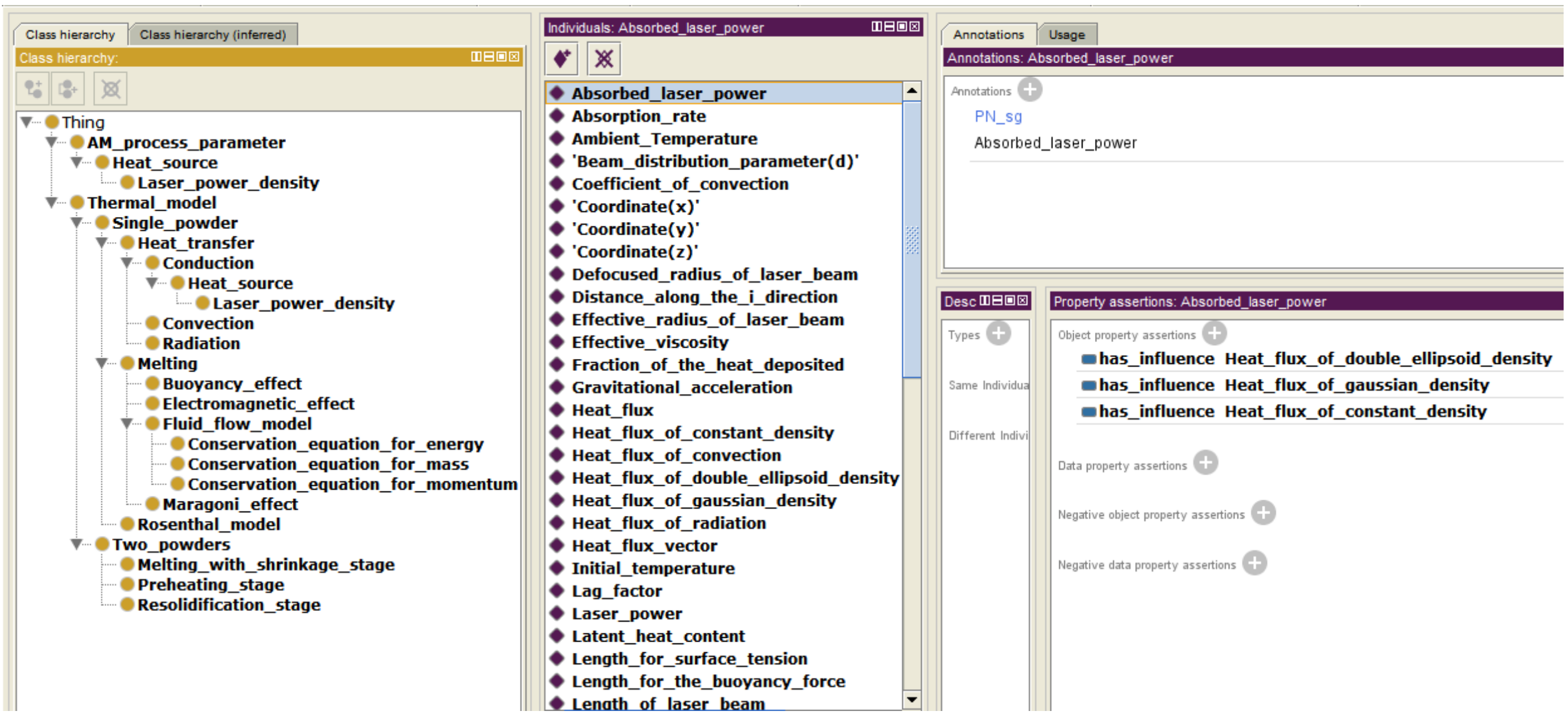

**Figure 16. Sample of AM ontology showing detail for absorbed laser power class definition**

The AM process ontology generates a network of parameters that can be visualized as a graph to look for similarities and differences across different models from different researchers. We will refer to these as ***knowledge graphs*** as they can be navigated forward (or backward) to identify important relationships between parameters and phenomenon that were previously disconnected. Two examples of navigating such a knowledge graph to identify important relationships during AM are shown in Figure 17. In Figure 17a, the knowledge graph is used to trace a process parameter that we can measure (i.e., melt pool area) to understand how it influences different mechanical properties that may be of interest (e.g., tensile strength, yield strength, elongation, Vickers hardness). The graph does not tell us exactly how they are related, but we know from the AM ontology that these parameters influence each other based on data and models in the literature.

These same ontologies developed to manage process models can be easily extended to support data management and configuration. As noted in Section III, a vast amount of AM process data is being measured, often used for development and validation of AM process models. Figure 17b shows an example of how the AM ontology can be leveraged to navigate the knowledge graph *in reverse* to identify what sensor data should be captured to help ensure that a requirement is met. In this example, we assume a requirement is specified on the Vickers hardness of the part, and then we navigate the knowledge graph backwards until we find measurable process parameters that we might be able to sense, namely, scanning speed and absorbed laser power in this case. While we may not be able to measure absorbed laser power directly, this nonetheless provides an indication of what we might want to sense during the process to gather data to help ensure that our requirement is met.

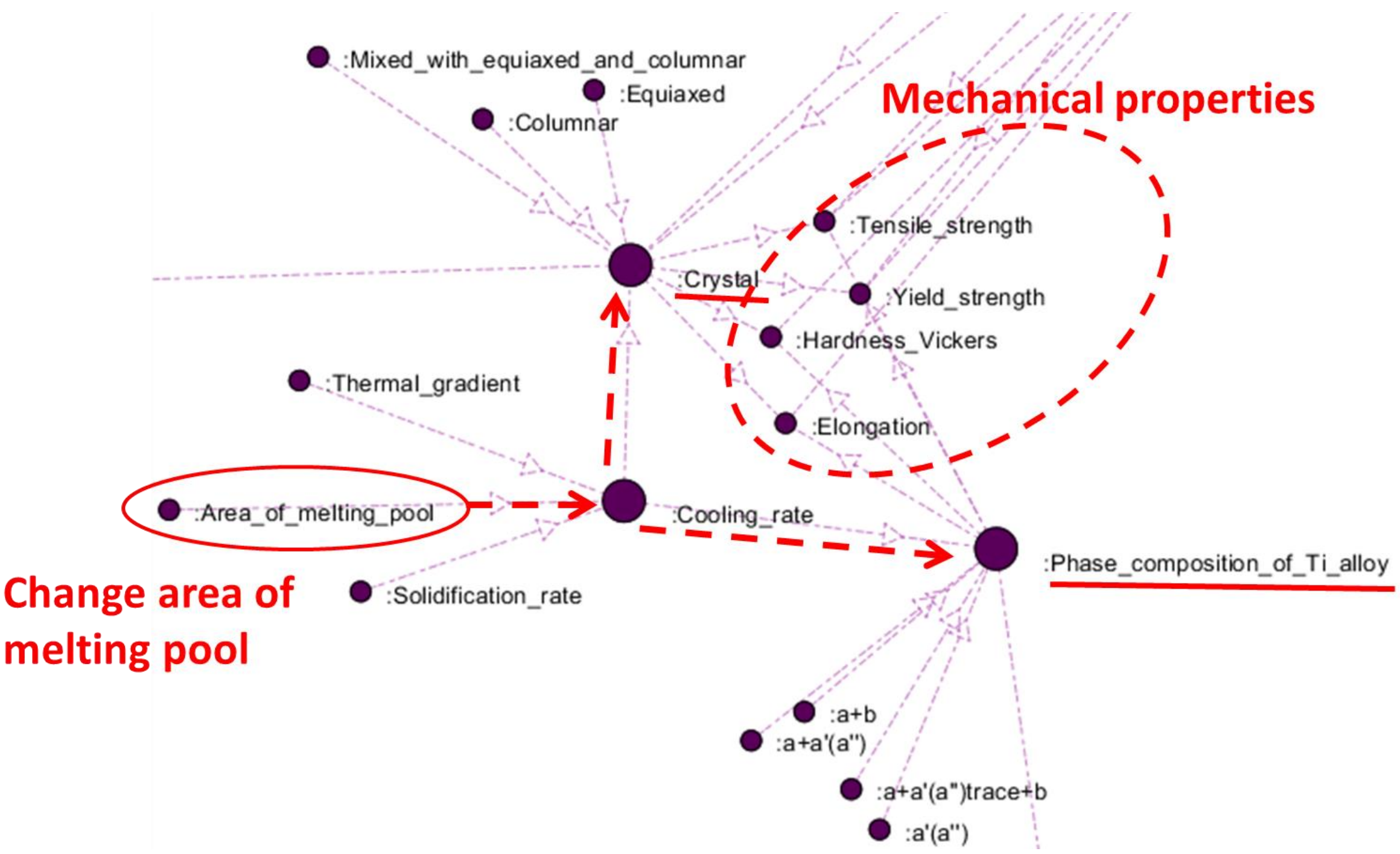

**(a) An example of using knowledge graph to navigate from a measurable process parameter (melt pool area) to mechanical properties of interest (tensile strength, yield strength, etc.)**

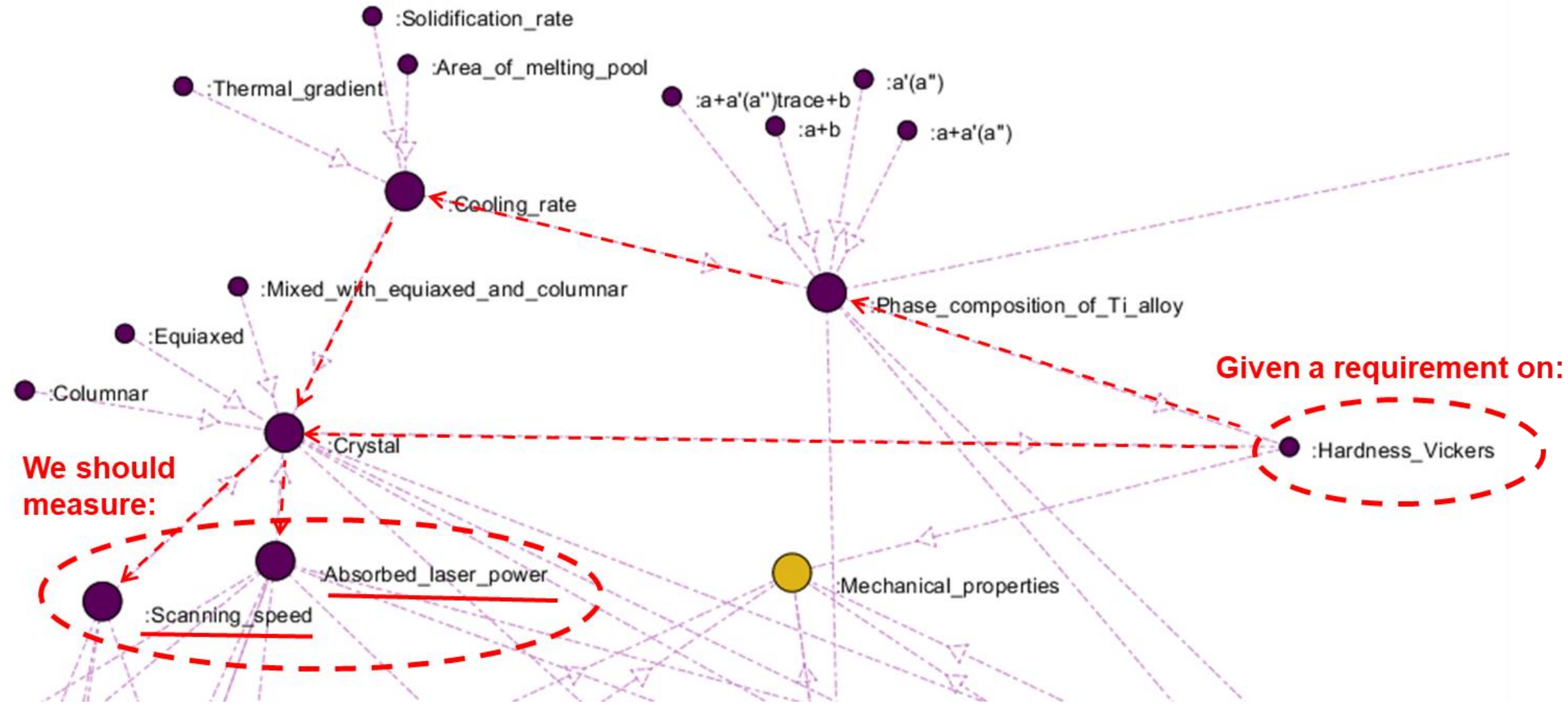

**(b) An example of navigating knowledge graph *backwards* to trace a requirement (Vickers hardness) to two measurable process parameters (scanning speed and absorbed laser power)**

**Figure 17. Examples of using knowledge graphs from AM ontology to identify relationships between measurable process parameters and potential requirements for a metal AM part**

The AM ontology and corresponding knowledge graphs can also be used to support analysis of process parameters and sensor data. For instance, Table 1 shows data from an experiment where several input process parameters (e.g., laser power, velocity, spot size) were varied, and sensors were used to capture melt pool depth and width; deposition height and width were also measured for each test specimen [80]. Linear regression was then used to analyze the data in Table 1, and $R^2_{adjusted}$ values for deposition height and deposition depth are 91.75 and 89.97, respectively, as a function of the process parameters that were varied. The $R^2_{adjusted}$ value for melt pool width is also good (94.85); however, the $R^2_{adjusted}$ value for melt pool depth is not (68.00). When we trace these relationships in our AM ontology, we find that the Marangoni effect, velocity of fluid, and Buoyancy effect have a relationship with melt pool depth, yet none of these are in the data *because they were not measured or sensed* during the experiment. Had the researchers had the ontology, the corresponding knowledge graph could have been used to plan the experiment more carefully, i.e., what data to sense and capture based on what they wanted to analyze after the experiment. This simple example demonstrates what might be achieved (and potentially avoided) by using a knowledge graph such as the AM ontology to guide sensing and inform analysis.

**Table 1. Experimental data for metal AM [80] and results of linear regression analysis**

| Sample No. | Power [W] | Velocity [in/min] | Velocity [mm/s] | Spot Size [mm] | Power Density [W/mm2] | Run time[sec] | Energy per mm [W/mm] | Melt-pool Depth | Melt-pool Width | Deposition Height | Deposition Width |
|---|---|---|---|---|---|---|---|---|---|---|---|
| Ti9 | 1000 | 10 | 4.23 | 1.89 | 318.31 | 34.116 | 236.22 | 86.5 | 2886.5 | 2108.1 | 4005.4 |
| Ti10 | 1000 | 10 | 4.23 | 2.89 | 79.58 | 34.116 | 236.22 | 75.7 | 3308.1 | 1800 | 4070.3 |
| Ti11 | 1000 | 10 | 4.23 | 3.89 | 35.37 | 34.116 | 236.22 | 75.7 | 2637.8 | 1935.1 | 4302.7 |
| Ti12 | 1000 | 25 | 10.58 | 1.89 | 318.31 | 13.6464 | 94.49 | 102.7 | 2508.1 | 437.8 | 2637.8 |
| Ti13 | 1000 | 25 | 10.58 | 2.89 | 79.58 | 13.6464 | 94.49 | 156.8 | 2378.4 | 637.8 | 2432.4 |
| Ti14 | 1000 | 25 | 10.58 | 3.89 | 35.37 | 13.6464 | 94.49 | 59.5 | 2400 | 756.8 | 3005.4 |
| Ti15 | 1000 | 40 | 16.93 | 1.89 | 318.31 | 8.59 | 59.06 | 162.2 | 2108.1 | 383.8 | 2394.6 |
| Ti16 | 1000 | 40 | 16.93 | 2.89 | 79.58 | 8.59 | 59.06 | 178.4 | 2059.5 | 616.2 | 2237.8 |
| Ti17 | 1000 | 40 | 16.93 | 3.89 | 35.37 | 8.59 | 59.06 | 59.5 | 1524.3 | 475.7 | 1735.1 |
| Ti18 | 2000 | 10 | 4.23 | 1.89 | 636.62 | 34.116 | 472.44 | 681.1 | 5005.4 | 864.9 | 5156.8 |
| Ti19 | 2000 | 10 | 4.23 | 2.89 | 159.15 | 34.116 | 472.44 | 140.5 | 4994.6 | 2064.9 | 5070.3 |
| Ti20 | 2000 | 10 | 4.23 | 3.89 | 70.74 | 34.116 | 472.44 | 97.3 | 4816.2 | 2108.1 | 5821.6 |
| Ti21 | 2000 | 25 | 10.58 | 1.89 | 636.62 | 13.6464 | 188.98 | 173 | 3902.7 | 1124.3 | 4010.8 |
| Ti22 | 2000 | 25 | 10.58 | 2.89 | 159.15 | 13.6464 | 188.98 | 291.9 | 3875.7 | 1135.1 | 3956.8 |
| Ti23 | 2000 | 25 | 10.58 | 3.89 | 70.74 | 13.6464 | 188.98 | 281.1 | 3881.1 | 724.3 | 4313.5 |
| Ti24 | 2000 | 40 | 16.93 | 1.89 | 636.62 | 8.529 | 118.11 | 389.2 | 3232.4 | 702.7 | 3491.9 |
| Ti25 | 2000 | 40 | 16.93 | 2.89 | 159.15 | 8.529 | 118.11 | 454.1 | 3351.4 | 518.9 | 3535.1 |
| Ti26 | 2000 | 40 | 16.93 | 3.89 | 70.74 | 8.529 | 118.11 | 324.3 | 3145.9 | 540.5 | 3340.5 |
| | *$R^2_{adjusted}$ values for linear regression against power, velocity, spot size:* | | | | | | | *68.00* | *94.85* | *91.75* | *89.97* |

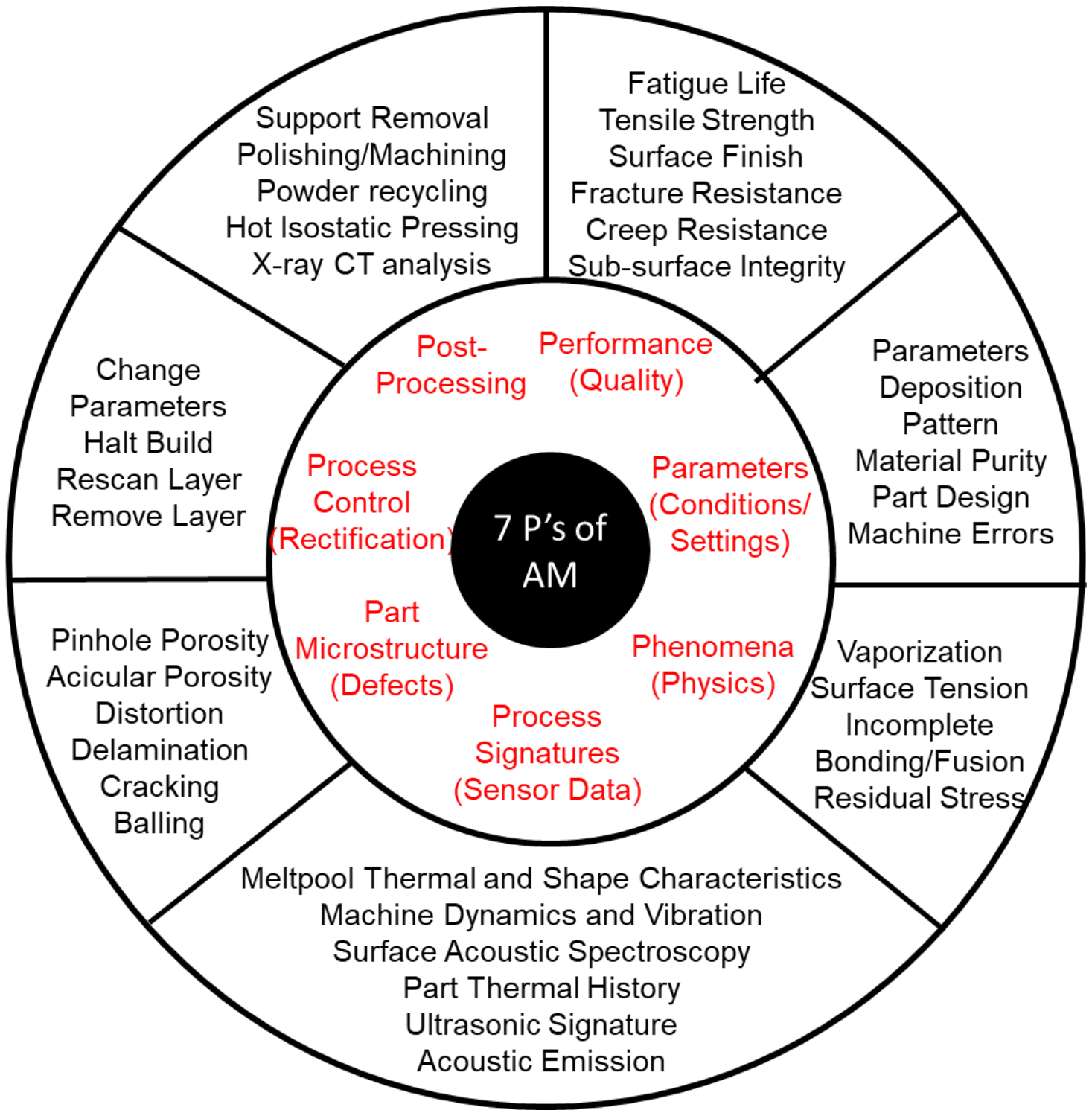

**Figure 18. The complex part design- process parameters – property linkages in additive manufacturing**

## B. Design of experiments (DOE)

The distinctive aspect of AM compared to traditional subtractive and formative manufacturing processes is the relative tight coupling of the part geometry (shape), microstructure evolved, and process conditions [81-83]. In other words, the shape, microstructure, and process conditions interact to influence the functional integrity aspects of the part, such as its strength, fatigue life, adherence to geometric and dimensional specifications, among others. This coupling of part shape, process parameters, microstructure, and part properties is rather weak in conventional manufacturing, for instance, in subtractive machining, although the near-subsurface microstructure is influenced by the cutting conditions and geometry, the bulk microstructure is largely unaltered. Some of these process-structure-property relationships in AM are exemplified in Figure 18.

This intricate interaction in AM lies at the crux of the large uncertainty in part quality aspects, and accordingly, the use of traditional DOE-based methods to achieve the optimal processing conditions is constrained for the following reasons:

*(1) A large number of key process input variables can be adjusted and several output variables need to be simultaneously optimized.*

For example, in the laser powder bed fusion process alone, a schematic of which is shown in Figure 19, over 50 process input variables are known to influence the part properties[73, 84]. Taking just the example of LPBF, the key input variables can be categorized into two main categories, namely, boundary condition factors, and input parameters as demarcated in Table 2. Within the former boundary conditions-related factors are two further divisions: (i) part design related, and (ii) material-related aspects. Under the category of controllable input factors, are three further sub-divisions, namely, (i) environmental factors; (ii) Process-machine factors, and (iii) the characteristics of the energy source, such as the laser, optics, and scanning factors.

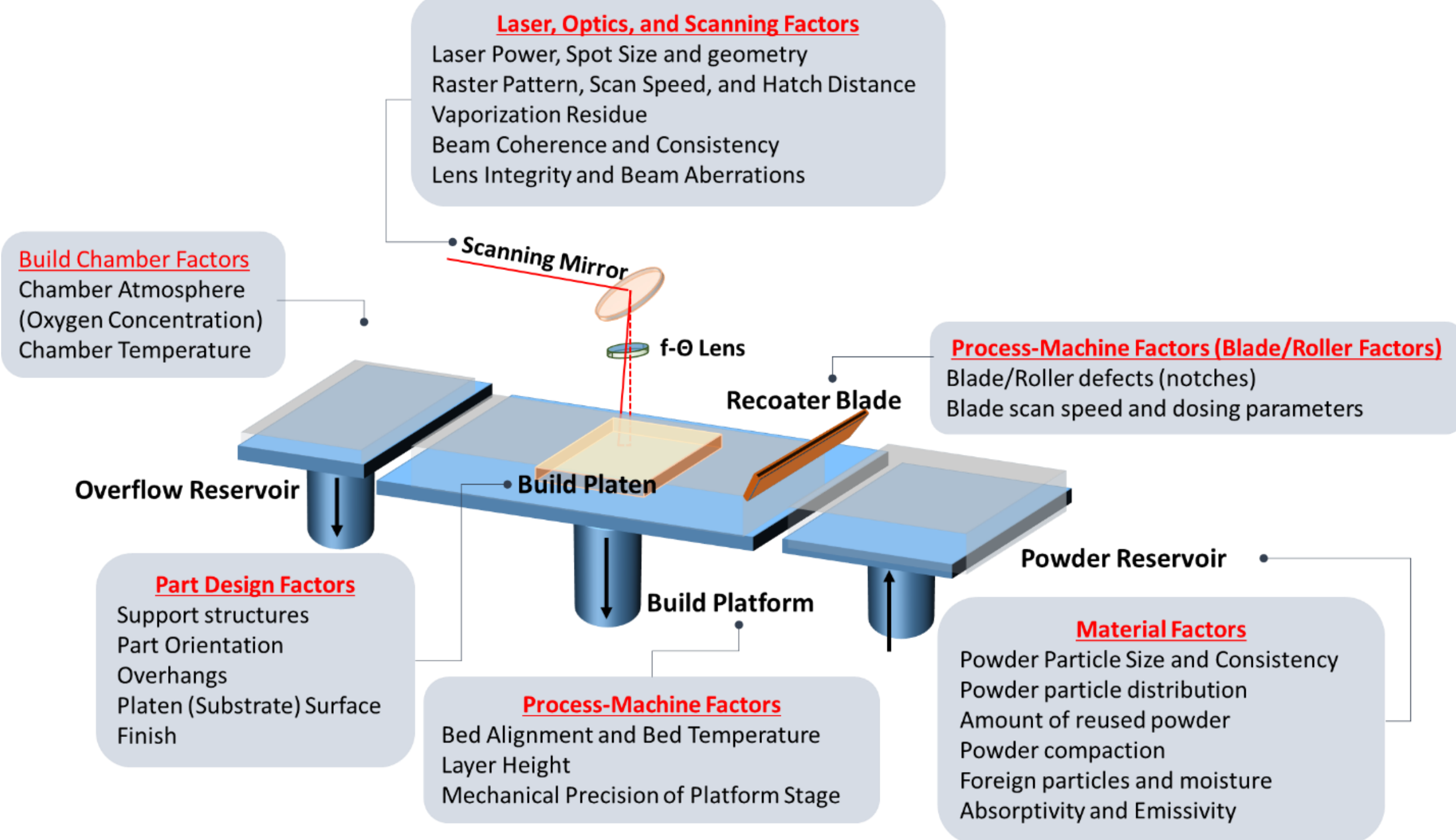

**Figure 19. The large number of process variables in LPBF additive manufacturing process makes process optimization using design of experiments expensive and untenable.**

Moreover, researchers have found that key process output variables may conflict with each other. For instance, part strength and geometric integrity are known to conflict; while increasing the infill percentage can increases the strength of the part, the increase in material density due to the added material has the tendency to create large residual stresses, that causes the part to warp [85, 86].

**Table 2. Boundary conditions and controllable input parameters in LPBF processes.**

| Boundary Condition Factors | | Controllable Input Parameters | | |
|---|---|---|---|---|
| Part and Build Factors | Material Factors | Environmental Factors | Process-Machine Factors | Energy Factors |
| • Location of support structures<br>• Contact area and type of supports<br>• Part orientation<br>• Overhang<br>• Platen (Substrate) thickness and finish<br>• Placement of parts on the bed | • Material type and purity<br>• Powder particle size and distribution<br>• Amount of powder reused from previous builds<br>• Powder compaction<br>• Foreign residue as a result of re-processing<br>• Absorptivity and Emissivity characteristics | • Oxygen concentration<br>• Chamber temperature<br>• Gas flow<br>• Spatter and debris<br>• Chamber evacuation gas (nitrogen, argon)<br>• Cleanliness of the lens and exhaust efficiency<br>• Presences of residue from previous builds | • Bed alignment (gap and skew)<br>• Bed temperature<br>• Layer height<br>• Precision of machine elements<br>• Blade type and rake angle<br>• Blade/Roller defects<br>• Blade scan speed and dosing parameters<br>• Interlayer cooling time. | • Laser, optics and scanning factors<br>• Laser power,<br>• Rastering pattern, scan speed, hatch distance<br>• Lens integrity<br>• Focus height above the bed. |

*(2) The influence of part geometry, process parameters, and build strategy on the build quality.*

In AM, mechanical and physical properties of the final part are governed by thermal aspects, such as heat flux and cooling time between layers. These thermal aspects are in turn functions of the process parameter, part geometry, support structures, and build plan. Hence, process parameters optimized for rudimentary test coupons established for one type of geometry may not typically carry over to another geometry. To explain further, currently, in metal AM processes, such as laser powder bed fusion (LPBF), process parameters such as the laser power (P) [W], velocity (V) [$m \cdot s^{-1}$], hatch spacing (H) [m], and layer height (T) [m] are aggregated in terms of the incident laser energy per unit volume, called global energy density, $E_v = P/(V \times H \times T)$ [$J \cdot mm^{-3}$], which when coupled with the scanning strategy determines the average rate of heat input at the build surface. However, the global energy density is not sufficient to ensure part quality, because, apart from the part geometry and process parameters, the placement of parts on the build plate, the shape and placement of other parts in the build plan (build layout) also influence the cooling rate. For example, Figure 20 depicts X-ray computed tomography images of the cross-section of an Inconel 718 cylinders made using the LPBF process [1]. The parts are built simultaneously using a commercial LPBF machine. The part demarcated as Disc B, is built under the so-called default, factory optimized process conditions recommended by the manufacturer for Inconel 718. Nonetheless, the part shows pronounced lack-of-fusion porosity.

Lack-of-fusion porosity, also called acicular porosity, occurs due to poor consolidation of the material with insufficient energy. The energy density for Disc B is close to 80 $J/mm^3$. However, increasing, indeed doubling, the energy density $E_v$, to 160 $J/mm^3$ as in the case of Disc A did not eliminate the lack-of-fusion porosity. The reason for this observation can be explained on the basis of the placement of the parts on the build plate, and requires an understanding of the manner in which the laser beam is focused on the powder bed. In LPBF systems, typically, the laser is in the infrared region with wavelength in the vicinity of 1050 nanometers, the beam is rastered with the galvanic mirror assembly in the x-y plane, and focused on the build plate by means of an optic called the f-θ lens. This lens is designed to maintain a constant focal length (f) irrespective of the angle of incidence (θ) of the laser beam after it is directed by the galvanic mirror assembly. A drawback with the f-θ lens is that at extreme incidence angles, corresponding to the edges of the build plate, the focal length tends to deviate from the desired setpoint. In other words, the beam tends to become defocused at the edges, and hence, building parts near the edges is not advisable, as the energy delivered will not be sufficient to melt the material. Some of the newer LPBF systems, such as the Renishaw RenAM 500M system have overcome this problem by replacing the f- θ lens with a dynamic focusing system.

We note that Both Disc B and Disc A are placed on the far corners of the build plate (the recoater scans from right to left), and since the LPBF system uses a f-θ lens, there is a possibility of exacerbated defocusing of the laser beam. This claim is substantiated in the case of Disc D, which has a smaller global energy density applied to it (107 $J/mm^3$) as opposed to Disc A, but is nominally devoid of porosity. This example serves to demonstrate that setting the process parameters to offline-optimized process conditions based on ideal conditions is not guaranteed to result in flaw-free parts in AM. Indeed, the placement of the parts on the build plate is also an important factor.

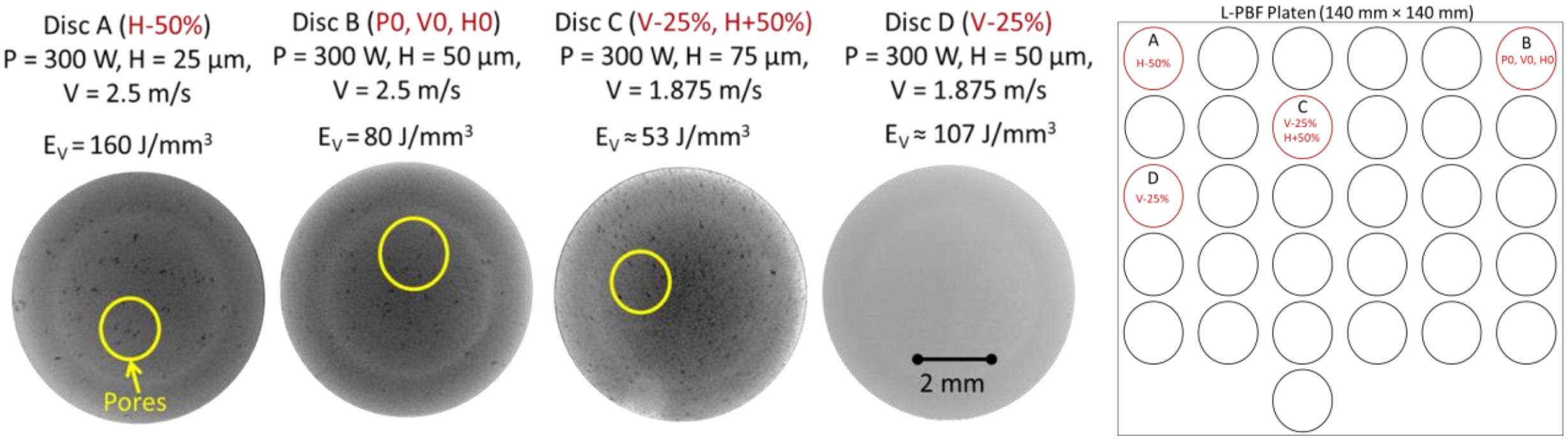

**Figure 20. The X-ray CT of the four discs, and their relative placement on the build platen [1].**

The geometry of the part beneath the powder bed in LPBF determines the rate at which the heat is conducted away from the build surface (heat flux), and hence governs the cooling rate, which in turn influences defects, such as cracking and microstructure heterogeneity. Placement of supports bares an important aspect of the part geometry because they serve as conduits for heat to dissipate [34, 87].

Furthermore, if more parts are added onto the build plate, the time for scanning a layer increases, and therefore, the heat from a previously melted region has a longer time to dissipate, which in turn alters the cooling rate. Thus, if any aspect of the build layout changes, for instance, new parts are added or taken away, the orientation of a component is altered, the scanning strategy and order is varied, then these changes will affect not just one part but potentially every part present on the platen during that build. Consequently, a part must be re-qualified when it is built as part of a different build layout.

*(3) Empirical testing is expensive.*

In AM, and more so in metal AM, the consumables are prohibitively expensive (the cost of powder material, such as titanium can exceed several hundred dollars per pound), the process is slow (a Φ8 mm × 60 mm tall build lasts close to 180 minutes), and only a few parts can be made at a time. Moreover, post-process destructive mechanical testing is expensive, and there are no standard approaches to ascertain the mechanical properties of complex objects, such as lattices. Indeed, non-destructive testing approaches, such as X-ray CT is cumbersome, and the resolution progressively degrades with the material density and size.

*(4) Sensitivity to disturbances (non-stationarity) makes maintaining stable experimental conditions difficult.*

One of the main tenets of statistical design of experiments is that the process should remain stationary during the duration of the test. This condition is not strictly true in AM, as the process parameters tend to fluctuate. For instance, in LPBF, during long experimental builds, the hot residue, such as vaporized material from the printing process tends to accumulate in the cooler areas of the machine. For instance, soot buildup on the optics leads to occlusion of the laser beam during long builds. Consequently, the shape of the laser beam and the power delivered tend to drift over time, which in turn affects the part properties.

Likewise, in directed energy deposition (DED) the morphology of the top surface of the part tends to change. In contrast to LPBF, the top surface in DED is not relatively flat but has an uneven wavy surface. This wavy surface originates because only a part of the material may be melted and adhered to the surface either due to a variety of reasons, such as insufficient energy to melt the surface, loss of powder in the stream, either too much or too less material flow. Subsequently, the distance between the top surface and the powder delivery nozzle (called the standoff distance) varies from its initial setpoint. If the standoff distance between the part and nozzle decreases, more power tends to be delivered, and accordingly more volume of the powder is melted, leading to a further decrease in the standoff distance. Eventually, the deviation of the standoff distance from the set point will rapidly exacerbate; the standoff distance will decrease and the nozzle may eventually crash into the part.

On the other hand, if the standoff distance increases, the power delivered is insufficient to melt the powder, and the standoff distance will decrease, causing the laser beam to *walk off* from the part. Such process drifts inherent to AM processes cause the part properties to vary along the build direction, and as a consequence induce a large spread in the measurement of the output variables.

For example, shown in Figure 21 is the X-ray computed tomography image of a titanium alloy coupon deposited using the DED process [88]. One of these parts which is deposited under sub-optimal process conditions; the laser power (300 W) is insufficient to melt the material, and manifests long lack-of-fusion flaws. When after extensive testing, it was found that when the laser power is increased from 300 W to 475 W, the lack-of-fusion flaws are mitigated, however a relatively small flaw is still evident, whose root cause cannot be pinpointed. In other words, there is a stochastic (random) aspect to defect formation.

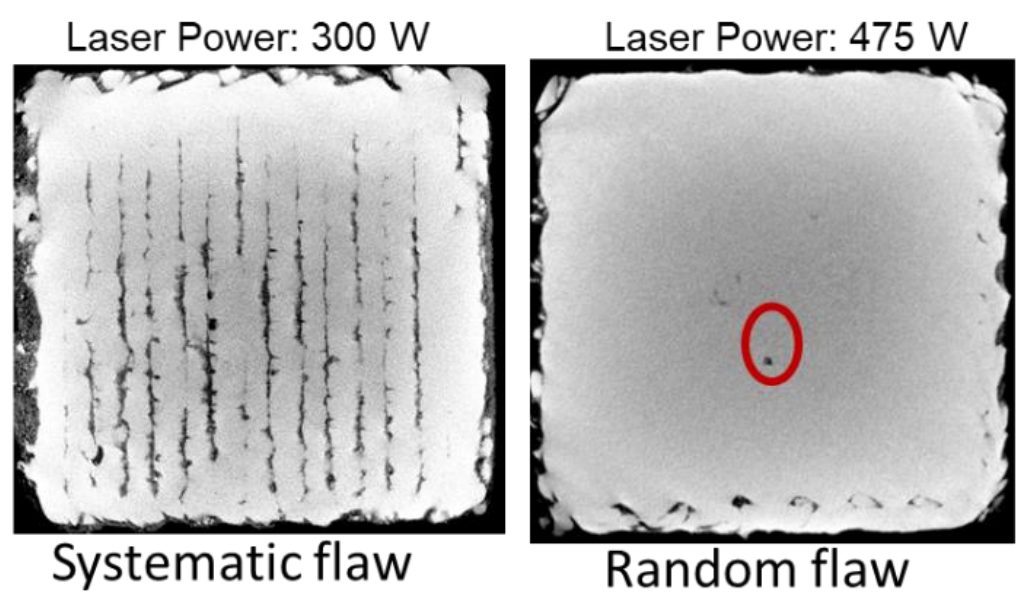

**Figure 21. Two DED parts (15 mm × 15 mm × 10 mm) show systemic flaws (left) due to poor selection of processing conditions, (right) random (stochastic) flaws tend to occur even under flaw-free conditions [88].**

These challenges pose considerable uncertainty in the generalizability and effectiveness concerning the conventional statistical design of experiments in AM. To address these concerns researchers have explored several strategies. First to reduce the number of expensive empirical tests required, sequential and evolutionary design of experiments strategies have been demonstrated [89]. The key idea of the evolutionary optimization approach is to use previous experiments to inform the next set of experiments. One approach to evolutionary optimization is to conduct a set of experiments, and test for the key process output variables. Based on the results, the next set of experiments are conducted in the vicinity of those process settings which result in outcomes closer to the desired. Another approach is to use a technique called minimum-energy design of experiments, which provides a set of candidate points using a Bayesian analysis [90].

Another strategy is to augment design of experiments with machine learning models trained on the available data set. In this regard, King *et al.* suggest including results from simulation models to rapidly narrow the process conditions. With regard to development of experimental data sets, extensive part design and testing strategies have been formalized by the ASTM F42 committee.

Note that that the global energy density is not sufficient to ensure part quality, because, apart from the part geometry and process parameters, the shape and placement of other parts in the build plan (build layout) also influences the cooling rate. The uncertainty introduced in the component quality due to the complex interdependence between material, part geometry, process parameters, and build plan negates one of the most attractive aspects of AM: the flexibility to implement changes to the part design without the need for extensive optimization of the process parameters. This process complexity in AM strengthens the case for supplanting an empirical build-and-test optimization approach for a thermal physics-driven methodology.

## C. Simulation modeling and analysis

Fast and accurate physical models are critically needed for AM to: (1) narrow the process parameter space for a property of interest, (2) identify red-flag problems in the part design, (3) aid support placement, build orientation, and build plans, (4) predict the distortion and microstructure evolved, and (5) augment process control by providing a model-based baseline for adjusting the process (feedforward control) [91-94]. From a broader vista, simulation in AM can be categorized into three classes contingent on the dominant phenomena, namely, thermal, fluid, and photopolymerization based. To explain further, in metal AM processes, such as LPBF and DED, the energy applied in joining the layers is supplied by a laser, accordingly, researchers have focused on modeling the thermal phenomena in metal AM. Melting and extrusion-based polymer AM approaches, such as fused filament fabrication may also be considered to fall under the category of thermal initiated AM. Processes such as binder jetting and aerosol jet printing are governed by the mechanics of droplet formation, fluid flow, and wetting. Lastly, material jetting stereolithography are governed by photochemical reactions.

In this paper, we have chosen to focus on metal AM processes given their popularity in high-value industries such as aerospace and biomedical. The industrial interests in LPBF and DED have propelled active research in simulation modeling of these processes, with several commercial ventures being initiated in the last decade. The three key problems faced by researchers in this area are as follows:

(1) Simulation time
(2) Coupling of phenomena across multiple scales
(3) Difficulty in experimental validation.

These difficulties originate because thermal modeling in LPBF and DED involves multi-scale physics, which starts at the meltpool level, progressing to the layer level, and finally the part level [39, 93, 95]. The various process-part thermal interactions in the LPBF and DED processes are depicted in Figure 22. The meltpool or particle-level dynamics are tied to material solidification rates, and the interaction of the laser beam with the powder, and hence is the key to predict the microstructure evolved, and as a consequence mechanical properties, such as hardness, strength, and fatigue-life [96]. Next, in ascending order, is the so-called meso-scale or track-level which ranges from a few hundreds of micrometers to under a millimeter. The aim of track-level simulations is to predict consolidation of the powder and dynamic evolution of the meltpool as the laser is scanned, which is consequential to the density of the part formed. Lastly, at the macroscopic- or part-level, which ranges from millimeters, and beyond, the thrust is to predict the thermal-related residual stresses and geometric deformation.

At the meltpool or particle-level, the interaction of the laser beam with particles is the focus. Particularly in LPBF, the energy absorbed by the material is a function of its reflectivity (in electron beam powder bed fusion the electronegativity is of importance). A highly reflective material will tend to absorb a smaller magnitude of the incident laser energy. Furthermore, the laser is reflected repeatedly by the powder particles when it is incident on the powder bed. This is advantageous to material melting as the energy absorbed by the material increases on account of multiple reflections. The laser-particle interaction is also important to understand the formation of pinhole (due to vaporization) and keyhole-type porosity. The former occurs due to one of three reasons. First, the vaporization of remnants of moisture on the surface of powder particles; second, the escaping gasses trapped within the meltpool; and third, due to the vaporization of impurities within the powder that have a lower melting point than the desired alloy. Keyhole melting porosity occurs at inordinately high laser energy conditions, which causes the powder to vaporize, and create a cavity. This cavity serves to further focus the laser into a narrow beam, exacerbating the vaporization of more material. Eventually, the surrounding material falls into the cavity and fills it incompletely, causing a void (keyhole collapse). In the case of the DED process, the simulation at this scale includes modeling the interaction of the falling powder with the carrier gas, as well as the laser.

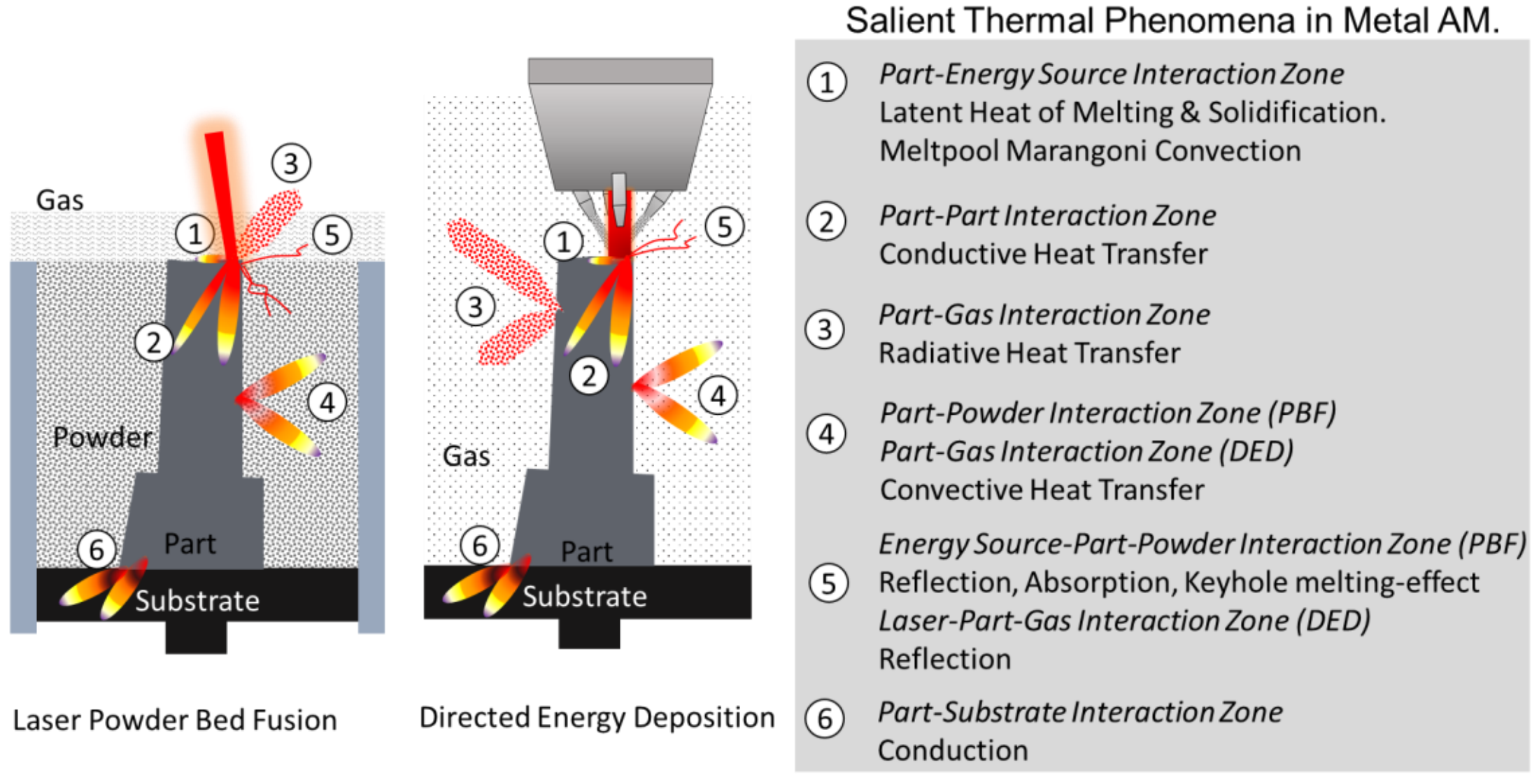

**Figure 22. The thermal phenomena in metal AM processes range across multiple scales, starting from the meltpool level to the part level [97].**

At the track-levels, the simulations must take into account surface tension-related phenomena, such as Plateau-Rayleigh effect which is the root cause of meltpool instability, and consequently, inferior consolidation of the material. Furthermore, at the meltpool-level the material changes from solid to liquid and back to solid again, as a result, latent heat effects cannot be neglected. The simulations at this scale have been used to model the segregation or breakup of the melptool into discrete chunks, called balling. This phenomenon is typically observed underneath of unsupported features in the part, and is related to the accumulation of heat in a region. The increase in temperature causes the surface tension of the meltpool decrease, which in turn leads to an increase in its length. The inordinate increase in the meltpool causes the onset of Plateau-Rayleigh instability causing the meltpool to break up into discrete chunks. Each of these chunks eventually coalesces into spheroid shapes. The occurrence of balling phenomena is tied closely to the laser power and hatch spacing. This example serves to emphasize that the dynamics of the meltpool and track-levels involve both fluid and heat transfer phenomena.

Lastly, at the Part-level prediction of the temperature distribution has garnered commercial interest, with the emphasis on four aspects: (1) predicting distortion during and after the build, (2) possibility of a recoater crash due to part distortion during the build, (3) optimizing part orientation and placement of supports, and (4) build layout planning. To explain further, the three main factors that influence the thermal distribution at the part-level in LPBF are:

(1) The geometry of the part, including features such as steep overhangs, and the presence of anchoring supports [87, 98-100].

(2) Type and characteristics of the feedstock material, and process parameters, such as the laser power, hatch spacing, layer thickness, laser scan velocity, and scanning strategy, which influence the average heat input (global energy density) [101].

(3) The time required for scanning a layer and the interval between the melting of successive layers (interlayer cooling time), which are functions of the build layout determined by the number, geometry, orientation, placement and scanning sequence of other parts on the build plate.

At the part-level, the effect of meltpool-level phenomena (e.g., latent heat aspects) are neglected to aid computation. Mathematically, the aim is to solve the heat diffusion equation, in which conduction is the model heat transfer, radiative and convective effects are considered post-facto, i.e., after the heat diffusion equation is solved. The heat diffusion equation takes the following form,

$$\rho c_p \frac{\partial \mathrm{T}}{\partial t} - k\left(\frac{\partial^2}{\partial x^2} + \frac{\partial^2}{\partial y^2} + \frac{\partial^2}{\partial z^2}\right)\mathrm{T} = \mathrm{Q}$$

Solving the heat equation results in the instantaneous temperature $\mathrm{T}(x, y, z, t)$ at a time $t$ for a Cartesian spatial coordinate (*x, y, z*). The temporal map of $\mathrm{T}(x, y, z, t)$, i.e., the trace of the temperature T at the location (*x, y, z*) over time, gives the temperature history in the part for that location. The right-hand side term is the energy supplied by the laser per unit volume of the material, per second (Q). Although the units are identical to the global energy density ($E_v$), Q is a more encompassing term because of the flexibility to include the effect of the beam shape.

The benchmark computational approach for solving the heat equation originates in the welding literature, exemplified in the work of Goldak *et al* [102]. This model called Goldak's double ellipsoid model, considers, as the name suggests, the laser source to be ellipsoidal in shape. The beam energy is assumed to be concentrated in the center, and dissipates near the boundary of the ellipse. In the AM context, researchers tend to consider the beam to be Gaussian shaped. A key difference between welding and AM is that the heat source has a smaller profile, and the translation speed (scan velocity) is magnitude higher. Consequently, the cooling rates in LPBF approaches the order of nearly a $10^5$ degrees per second. In DED, the spot sizes are much larger than LPBF.

The main problem faced at the part-level thermal modeling is the evolving nature of the part geometry in AM. To explain further, the model must take into account the change in the computational domain and boundary conditions as material is deposited layer-upon-layer. The key challenge is to keep track of the elements from a finite element modeling perspective [103]. Typically, this is done through the element birth-and-death approach, where elements are slowly activated. The second is the quiet element method, wherein the part is meshed beforehand but the thermal properties of an element are activated at the appropriate interval. Commercial software, such as Netfabb, make use of a hybrid strategy involving both the quite element and birth-and-death approach. It may be noted that researchers at Lawrence Livermore National Laboratories have developed comprehensive multi-scale modeling tools based on their extensive code base, at the meso-scale (ALE3D) to the part-level (Diablo). Techniques such as finite difference and discrete element methods have been employed to solve the heat diffusion equation [104]. Newer approaches based on circuit theory and graph analysis have been introduced for mapping the thermal distribution in AM [97, 105].

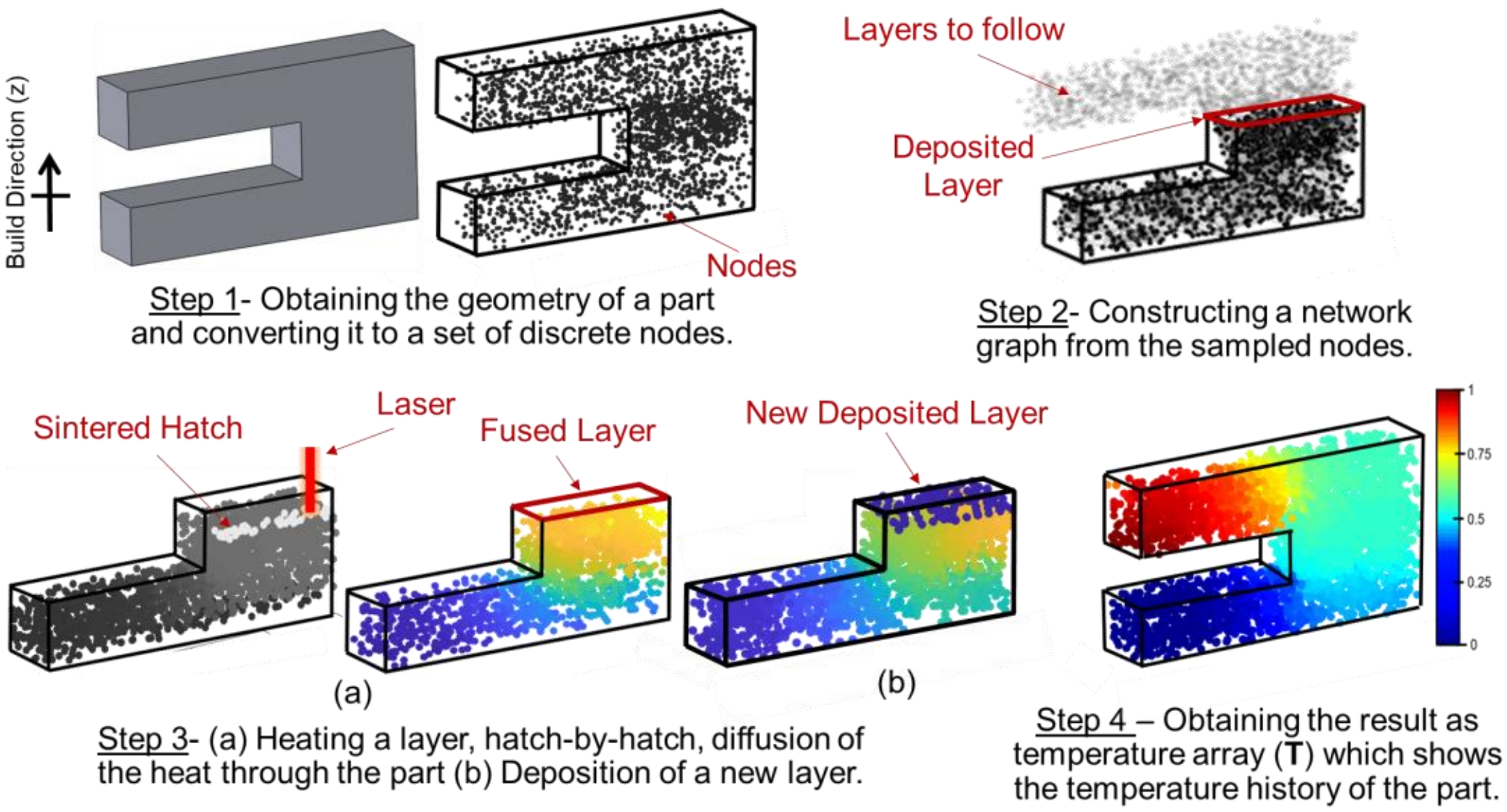

**Figure 23. The graph theory approach for simulating the LPBF process [97, 105].**

Figure 23 shows a schematic of the mesh-free graph theory to solve the heat diffusion equation. The key idea is that the discrete heat diffusion equation is solved as a function of the eigenvalues and eigenvectors of the Laplacian matrix of a graph projected onto the geometry of the part. The main advantage of the graph theory approach is that the temperature distribution part can be potentially computed many times faster compared to FE because: (1) graph theory eliminates time-consuming meshing steps and (2) it avoids cumbersome matrix inversion operations needed to solve the heat equation, and instead uses matrix Eigen decomposition.

Further, the graph theory approach is verified with a finite element implementation of the so-called gold standard Goldak's double ellipsoid model, which has its genesis in welding [102]. The graph theory solution was also quantitatively compared with the commercial Netfabb solution. The results for three-part geometries are shown in Figure 24. The graph theory simulation accurately predicts the accumulation of heat in the overhang region of a C-shaped part. Moreover, the approach also predicts that heat trapped in an overhang region can be dissipated by build extra supports. More pertinently, at the graph theory approach converged to within 90% of Goldak's solution within 10% of the computation time. The fast convergence of the graph theory approach opens the possibility for recognizing and correcting red flag problems in part design even before the part is printed. In other words, thermal simulations can be used as a viable path for design optimization in AM.

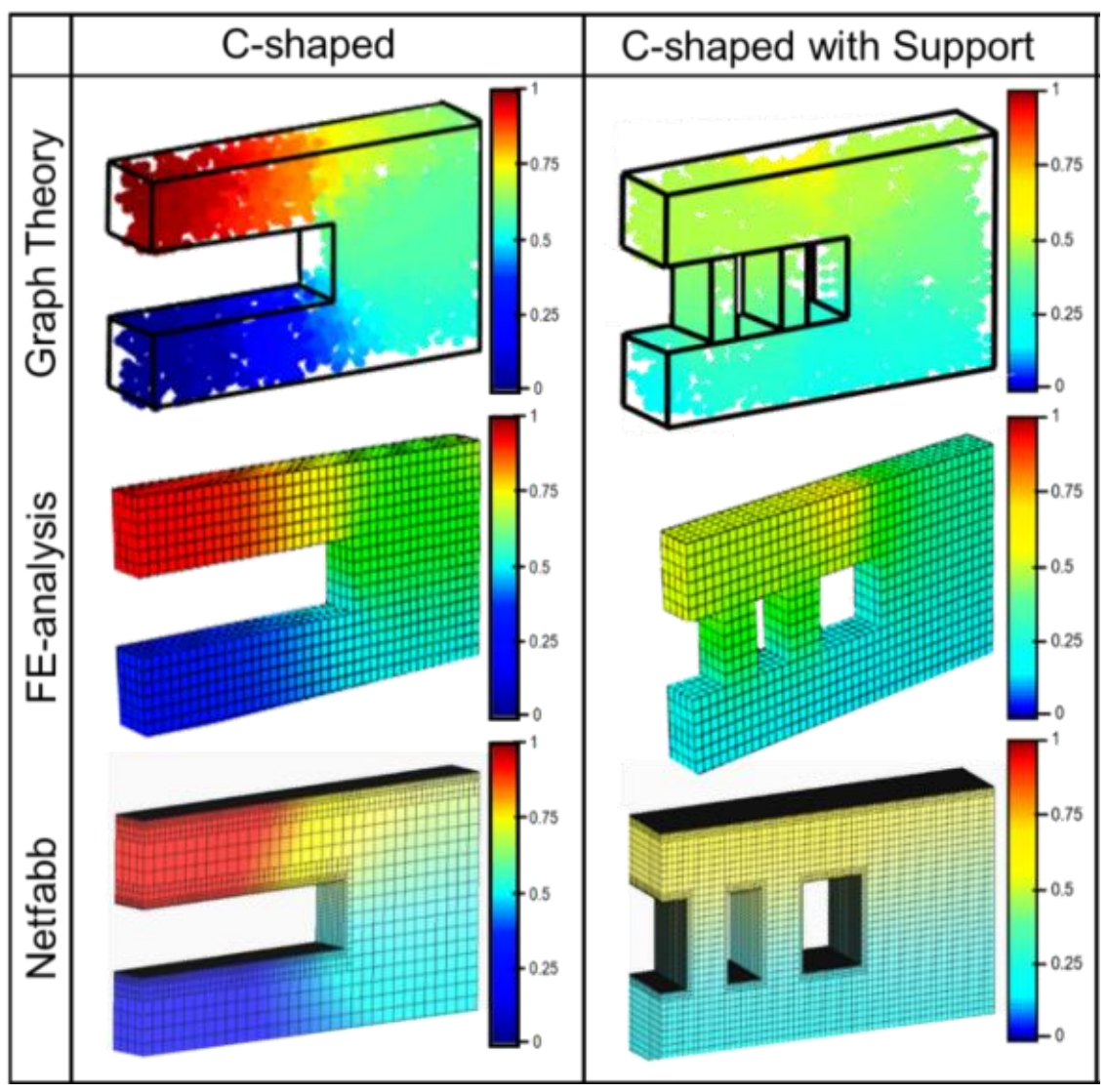

**Figure 24. Comparison of the graph theory approach with an FE-implementation of Goldak's model and the commercial Netfabb software for three different part geometries [97, 105]. The images are the temperature distribution in the last layer of the part (the part is 20 mm long × 2 mm wide × 20 mm tall). The temperature distribution is shown in normalized units.**

# VI. Control the Process

This section presents the learning and optimization of action strategies for AM quality control when the state of the build is dynamically evolving from one layer to another. As the finish in each layer will impact the next layer and all subsequent layers, this is a typical sequential decision-making program under real-world uncertainty (e.g., random variations, perturbations, or errors from measurements, machine settings, environments, and statistical estimation). Further, we present a constrained framework for sequential decision making. Examples of constraints include the lead time to complete a build, materials and/or energy consumption in the manufacturing process.

## A. Sequential decision making under uncertainty

Modern industries pose more stringent standards in product aesthetics, quality assurance, and functional integrity. Thus, it is critical for AM machines to increase the capability to mitigate incipient defects. Hybrid machines with both additive and subtractive manufacturing abilities provide an opportunity to make layerwise repairs, take corrective actions, thereby realizing a new paradigm of zero-defect AM [106]. For instance, sensor-based analytical methods (also see Section IV) help characterize and estimate the state of defects in each layer of the AM build. If a layer is estimated to have a small likelihood $s_l$ to contain defects, the AM process will continue and take no corrective action, denoted as $a_W$. On the other hand, if a layer has a high likelihood $s_h$ to have embedded defects, the AM process will pause and take an action to machine off this defective layer, denoted as $a_M$. The number of available actions depends to a great extent on the technological advancement of hybrid machines. For example, for the defects due to lack of fusion, a potential action is to re-fuse with the laser and mitigate such defects, denoted as $a_L$. The more actions available, new sequential optimization methods are more urgently needed for AM. This is mainly due to the fact that AM layers are not independent but rather highly correlated with each other. As shown in Figure 25 (a), the action chosen for one layer will impact the evolving dynamics of defect states in the next layer and, through that, all subsequent layers. In addition, there are uncertainties in the sensor measurements, machine settings, environments, defect estimation, and layer-to-layer transitions. The new sequential optimization framework needs to account for the uncertainty in AM processes and realize the zero-defect AM by minimizing the expected cumulative cost at the end when all layers are completed.

As shown in Figure 25 (b), each layer of AM builds will be captured by the sensors (i.e., high-resolution cameras) as imaging profiles. The probability for a layer to contain defects (e.g., $s_h$, $s_m$, $s_l$) will be estimated with sensor-based analytical methods such as layerwise deep learning of incipient defects [74, 75]. The *sequential decision-making framework for smart AM is formulated through a Markov Decision Process (MDP) model*. Although the MDP has been widely used and proven to be effective in the management of engineering systems [107-110], very little has been done to realize smart AM using MDPs. Our prior work formulated this problem as a Markov decision process corresponding to a 5-tuple $(\Omega, S, A, \mathcal{T}, R)$, where $\Omega$ is the set of sensor observations; $S$ is the set of defective states; $A$ is the set of actions; $\mathcal{T}: S \times A \times S$ represents the state transition; and $R$ is the reward function. The main objective is to search for an optimal policy $\pi^*(s)$ specifying the optimal action $a^*$ in state $s$, which will maximize the sum of rewards after taking the action $a^*$ and thereafter keeping being optimal.

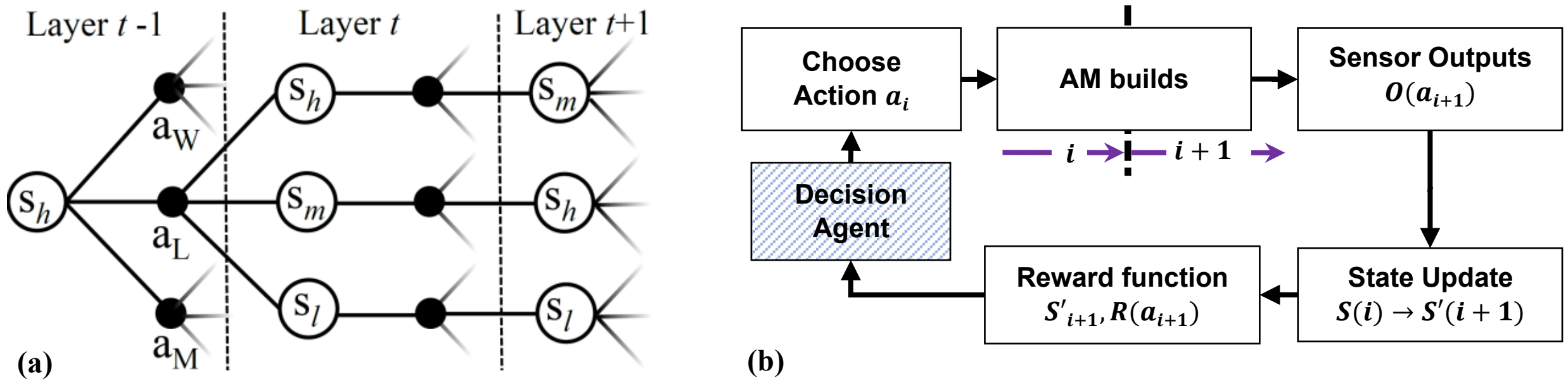

**Figure 25. (a) An illustration of state-action transition diagram. Note that $s_h$, $s_m$, and $s_l$ denote the high, median, and low defect states of an AM layer; (b) Markov decision process for smart AM.**

(a) *States, Actions, and Observations*: The complexity of AM poses challenges in measuring and characterizing the exact defective state of a layerwise build. As shown in Figure 14, we have developed a DNN learning method that tackles the challenge of layerwise geometrical variations and then estimate the risk probability of defects in a layer. As such, we can take full advantage of in-process image profiles and integrate them with MDP models. Each action affects the state transitions between layerwise builds in the AM process. Here, actions that are generally available in hybrid AM may include do nothing, cutting off a layer, re-fusion, or process adjustments.

(b) *State Transitions*: $p(s, a, s')$ provides the probability that an intervention $a$ in state $s$ at layer $i$ will lead to the state $s'$ at layer $i + 1$. The transition can be estimated from rich data collected in AM processes, but is influenced by the uncertainty in sensor measurements and process conditions. Few works in the AM literature studied sequential decision making under uncertainty.

(c) *Reward Function*: $R(s, a, s')$ is a reward that the decision agent receives for a specific state transition. For example, if an action drives the defect likelihood from high to low, it will be rewarded. Otherwise this action will be penalized. The utility $V^*(s)$ represents the sum of rewards received when starting in the state $s$ and acting optimally, and $Q^*(s, a)$ is the utility when taking the action $a$ from the state $s$ and thereafter acting optimally.

Further, we performed preliminary studies to develop a novel "sensing-modeling-optimization" framework that is tailored for AM processes. First, we leveraged the advanced sensing capabilities readily available in Penn State CIMP-3D to collect large amounts of layerwise image data. Second, we developed new sensor-based models to estimate the risk probability for a layer to contain defects, then predict the evolving dynamics of defect conditions from one layer to the next. Third, new MDP models are developed to model state-action transition dynamics among layers as a stochastic Markov process, and further derive the optimal control policy [111, 112]. The new "sensing-modeling-optimization" framework enables the implementation of in-process corrective actions to repair and counteract incipient defects in each layer of AM builds prior to the completion. The propagation of defects will be detected by sensor-based modeling and analysis of in-situ data, and will be mitigated long before they reach the non-recoverable stage.

## B. Constrained optimization of AM processes

Markov decision process (MDP) helps optimize the policy to choose layerwise actions by maximizing expected rewards (or minimizing the expected cumulative cost incurred by the AM defects) for a sequential decision-making problem in the real-world AM environment. Traditional MDP frameworks commonly focus on a single objective (e.g., minimizing the defects in each layer of AM build) [111], and is less concerned about multiple simultaneous objectives that may be added on the AM processes (i.e., minimizing total cost – wasted materials, consumed energy, or lead time, as well as improving the quality). As a vertical step to advance smart and sustainable AM, there is an urgent need to investigate multi-objective optimization of sequential decision-making problem for Six-Sigma quality management of AM.

If there are multiple objectives, e.g., minimizing total cost (e.g., lead time or consumed energy) in the AM process while improving the quality of layerwise builds, then sequential optimization becomes a challenging task because some objectives may be conflicting with others. For instance, if we increase the frequency to take corrective actions and make sure each layer has a small likelihood to contain defects, then the lead time to complete the build will be longer and more energy will be consumed. In other words, the number of defects will be minimized in each layer of the build but the total cost will be high. On the other hand, if we do not take as many corrective actions as needed, the build can be completed in a shorter period of lead time and less energy will be consumed. The total cost is low, but the likelihood to contain defects in the AM build will be higher. In the state of the art, few, if any, previous works have considered multi-objective optimization of the sequential decision-making strategy for AM processes. In particular, there is a need to balance multiple conflicting objectives for the quality management of AM builds.

To address these challenges, our prior work proposed a new constrained MDP (CMDP) framework to derive the optimal control policy in each layer of the AM processes that minimizes the total cost (e.g., lead time or consumed energy) and makes sure the quality standards are met for the AM builds [112]. The CMDP formulation is detailed as follows:

**State space**: The state space is defined as $\mathcal{S} = (\boldsymbol{T}, S)$, where $\boldsymbol{T} = \{1,2, \cdots T\}$ denotes the set of layer index, and $S$ is the set of defect states, i.e., $s_1, s_2, \cdots s_l$, which is structured in the increasing order of defect levels (i.e., $s_1$ is the lowest defect level, and $s_l$ denotes the highest defect level).

**Action space**: In the present study, the action space is simplified to include three actions, $A = \{a_M, a_L, a_W\}$, where $a_M$ denotes the action of removing a layer with the cost of $c_M$, $a_L$ is the action of laser repair and re-fusion with the cost of $c_L$, and $a_W$ represents the action to do nothing with the cost of $c_W$. With rapid advances of hybrid AM technology, it is anticipated that more actions will be available with different costs to be considered in the future work.

**Decision policy**: Let $Q_t(s_t, a)$ denote the decision rule at layer $t$, which is defined as the probability to choose an action $a \in A$ given the presence of defect state $s_t$ at the layer $t$.

**State transition**: Let $P_t^a(s_{t+1}|s_t)$ be the transition probability from state $s_t$ of layer $t$ to state $s_{t+1}$ of layer $t+1$ under the action $a \in A$. Given the decision policy $Q_t(s_t, a)$, the state transition is then defined as

$$M_t(i,j) = \sum_{a \epsilon A} Q_t(s_t, a) P_t^a(s_{t+1} = s_j | s_t = s_i)$$

Let the vector $\boldsymbol{x_t} = [x_{t1}, \ldots, x_{tl}]^T$ ($\boldsymbol{1^T x_t} = 1$, where $\boldsymbol{1}$ is a vector of 1's) represent the probability distribution of defect states $s_t \in \{s_1, \ldots, s_l\}$ at layer $t$, which means the probability of defect state $s_t$ staying in the defect level $s_i$ is $x_{ti}$. Then, $\boldsymbol{x_t}$ evolves according to:

$$\boldsymbol{x_{t+1}} = \boldsymbol{M_t x_t}$$

The CMDP model will then be formulated as follows:

$$\min_{Q_1, \ldots, Q_{T-1}} v_T = E_{x1}\left[\sum_{t=1}^{T-1} c_t(x_t, Q_t) + c_T\right]$$

$$s.t. \quad \boldsymbol{x_t} \leq \boldsymbol{h}, \boldsymbol{1}^T \boldsymbol{x_t} = 1$$

$$\boldsymbol{x_{t+1}} = \boldsymbol{M_t x_t},\ Q_t \boldsymbol{1} = 1, Q_t \geq 0$$

$$\text{for } t = 1,2, \ldots, T-1$$

where $Q_t$ is the decision matrix for layer t, $v_T$ is the expected total cost in energy or time, $c_t(x_t, Q_t) = \sum_{a \epsilon A} c_a Q_t(s_t, a)$ is the immediate cost at layer t, and $c_T$ is the terminal cost at the final layer $T$. The first constraint makes sure that the quality standards are met by bounding the probability of each defect state with an upper limit $\boldsymbol{h}$ and $0 \leq \boldsymbol{h} \leq 1$. The second constraint guarantees each row of $Q_t$ to be a valid probability distribution. If we delete the quality constraint (i.e., $\boldsymbol{x_t} \leq \boldsymbol{h}$) in the CMDP model, then the rows of $Q_t$ will be independent and not correlated. As such, the CMDP model can be solved with dynamic programming and simple backward induction algorithms. However, due to the quality constraint on the density distribution $\boldsymbol{x_t}$ of defect state $s_t$, the rows of $Q_t$ are correlated in the formulation through state-action transition dynamics $\boldsymbol{x_{t+1}} = \boldsymbol{M_t x_t}$. As a result, it is difficult to solve the CMDP model here with traditional dynamic programming algorithms. Therefore, our prior work developed new dynamic programming algorithms to solve the CMDP model and demonstrated the optimal control policy for the worst-case scenario of probability distribution of defect states [112].

In the proposed "sensing-modeling-optimization" framework, in-situ sensor signals, which exemplify specific process defects, are integrated with AI, machine learning, and CMDP models to optimize the selection of corrective actions for smart and sustainable AM. In addition, the objectives can also be extended to include the minimization of delamination and warpage of the final workpiece, and the maximization of reliability measures such as build strength and fatigue resistance. As opposed to purely data-driven approaches, which cannot suggest process adjustments, this sensor-based modeling and optimization approach not only detects process anomalies, but also guides the optimal corrective action, thereby enabling close-loop control of AM build quality and functional integrity.

# VII. Discussion and Conclusions

AM provides an unprecedented opportunity to produce complex geometries that are often impossible with traditional subtractive (machining) and formative (casting, welding, molding) manufacturing processes. Once the quality challenge is tackled, such a capability will result in the advent of newer and cheaper consumer products. Also, AM offers the possibility of taking a computer-generated design and directly putting the build into the hands of an end-user. If designs can be repeatably produced with a very low probability for defects, then new disruptive business models will become possible. A brick-and-mortar retail store will no longer need to carry an inventory of final products. A consumer could simply go to the store or the store's online website, select a premium and validated product from a catalog, push a button, and wait for the product to be made using an AM process. This so-called "zero lead time" store could see extended applications with at-home AM machinery and systems. Digital designs could be downloaded from the internet and created in the comfort of one's own home. Nonetheless, these concepts of "zero defects" or "zero lead time" depend to a great extent on the effective management of AM quality to recognize and anticipate defects and then take the appropriate corrective action to control process variability and ensure the final build's conformance to standards.

However, effective management of AM quality cannot just rely on the purchase of new machines and the installation of sensing and automation systems, but rather requires a set of quality-focused activities, ranging from quality planning, QA/QC, and continuous quality improvement. Quality planning identifies the needs of AM customers, e.g., whether they are interested in zero-defect products, aesthetic aspects, or geometric accuracies. Only by listening to the customers, can AM manufacturers develop the right strategic plan to help save time and costs in the handling of product returns, warranty charges, customer complaints. QA/QC focuses on the reduction of process variability and ensures the quality levels of final builds meet with standards (or specifications) from the customers. An important QA/QC function is to develop the ontological knowledge graph, document fundamental elements of the AM process (e.g., suppliers, materials, machines, processes, outputs, and customers in the AM ontology), analyze their relevance to the product quality, and identify the responsibilities (and accountability) of each element or business unit. Quality improvement goes beyond QA/QC activities to engage in the continuous improvement of quality towards gaining competitive advantages in the global market. As mentioned in Section V, ontology analytics, design of experiments, and simulation analysis are major methods and tools that can be used to help AM manufacturers further improve quality in a continual basis.

Furthermore, quality management is not just the job of quality-inspection unit in an AM enterprise, but rather depends on all units during the AM process. For example, the design should consider the capability of AM machines and then be optimized for quality. The selection of suppliers should not be based on the cost only, but also focus on the quality and timely delivery of raw materials etc. Indeed, quality management should include engineering, operational, and managerial activities to ensure that the AM builds are conforming to standards and then continuously engage in the quality improvement. On the other hand, quality should not become nobody's job once everybody is involved. QA/QC are needed to develop the documentation and policy to explicitly provide the quality-related responsibility and accountability of each person or business unit during the AM process, from procurement engineers to machine operators to higher levels of management and so on. The philosophy of quality management is to emphasize on quality, raise the awareness, and engage each person in the AM process, then communicate quality problems effectively so as to optimize resource allocation and tackle such problems efficiently.

Lest quality-related challenges with AM are addressed, it is unlikely that traditional manufacturers will forego well-established conventional methods. In light of the strategic and economic prize at stake, there is a burgeoning need to address the quality challenges in AM, reduce process variability, and improve AM process repeatability. This paper aims to advance the scientific basis of AM quality management. The DMAIC approach for AM quality improvement has the potential to substantially improve production-scale viability of AM and enable wider exploitation of AM capabilities beyond the current rapid prototyping status quo. Achieving quality excellence in AM may have consequential socioeconomic impacts and outcomes, in terms of profitability (quick scaling of process conditions to changing requirements), sustainability (economy of resources and energy by reduction in waste, scrap and rework), and efficiency (minimize efforts required towards obtaining the best quality product). This will spur the growth of advanced manufacturing in the nation and the world, thus leading to broader social and economic impacts. This paper will help catalyze more in-depth investigations and multi-disciplinary research efforts to advance the new practice of Six-Sigma quality management for AM.

# References

[1] M. Montazeri, A. R. Nassar, A. J. Dunbar and P. Rao, "In-Process Monitoring of Porosity in Additive Manufacturing using Optical Emission Spectroscopy," *IISE Transactions,* pp. 1-28, 2019.

[2] T. Dasgupta, "Using the six-sigma metric to measure and improve the performance of a supply chain," *Total Quality Management & Business Excellence,* vol. 14, no. 3, pp. 355-366, 2003.

[3] G. E. P. Box and W. H. Woodall, "Innovation, Quality Engineering, and Statistics," *Quality Engineering,* vol. 24, no. 1, pp. 20-29, 2012.

[4] M. Seifi, M. Gorelik, J. Waller, N. Hrabe, N. Shamsaei, S. Daniewicz and J. J. Lewandowski, "Progress Towards Metal Additive Manufacturing Standardization to Support Qualification and Certification," *Journal of the Minerals Metals and Materials Society,* vol. 69, no. 3, pp. 439-455, 2017.

[5] K. Jurrens, "Measurement Science Roadmap for Metal-Based Additive Manufacturing - workshop summary report," *National Institute of Standards and Technology,* pp. 1-86, 2013.

[6] E. Sachs, M. Cima, J. Cornie, D. Brancazio, J. Bredt, A. Curodeau, T. Fan, S. Khanuja, A. Lauder, J. Lee and S. Michaels, "Three-Dimensional Printing: The Physics and Implications of Additive Manufacturing," *CIRP Annals,* vol. 42, no. 1, pp. 257-260, 1993.

[7] D. Pal, N. Patil, K. Zeng and B. Stucker, "An Integrated Approach to Additive Manufacturing Simulations Using Physics Based, Coupled Multiscale Process Modeling," *Journal of Manufacturing Science and Engineering,* vol. 136, no. 6, pp. 061022-061022-16, 2014.

[8] W. E. King, A. T. Anderson, R. M. Ferencz, N. E. Hodge, C. Kamath, S. A. Khairallah and A. M. Rubenchik, "Laser powder bed fusion additive manufacturing of metals; physics, computational, and materials challenges," *Applied Physics Reviews,* vol. 2, no. 4, pp. 041304, 2015.

[9] M. Matthews, J. Trapp, G. Guss and A. Rubenchik, "Direct measurements of laser absorptivity during metal melt pool formation associated with powder bed fusion additive manufacturing processes," *J. Laser Appl.,* vol. 30, no. 3, pp. 032302, 2018.

[10] C. B. Stutzman, A. R. Nassar and E. W. Reutzel, "Multi-sensor investigations of optical emissions and their relations to directed energy deposition processes and quality," *Additive Manufacturing,* vol. 21, pp. 333-339, 2018.

[11] E. W. Reutzel and A. R. Nassar, "A survey of sensing and control systems for machine and process monitoring of directed-energy, metal-based additive manufacturing," *Rapid Prototyping Journal,* vol. 21, no. 2, pp. 159-167, 2015.

[12] B. Foster, E. Reutzel, A. Nassar, B. Hall, S. Brown and C. Dickman, "Optical, layerwise monitoring of powder bed fusion," in *Solid Freeform Fabrication Symposium,* Austin, TX, 2015, pp. 295-307.

[13] A. R. Nassara, E. W. Reutzel, S. W. Browna, J. P. Morgana Jr, J. P. Morgana, D. J. Natalea, R. L. Tutwilera, D. P. Fecka and J. C. Banksa, "Sensing for directed energy deposition and powder bed fusion additive manufacturing at penn state university," in *Laser 3D Manufacturing III,* San Francisco, CA, 2016, pp. 97380R.

[14] C. Gobert, E. W. Reutzel, J. Petrich, A. R. Nassar and S. Phoha, "Application of supervised machine learning for defect detection during metallic powder bed fusion additive manufacturing using high resolution imaging." *Additive Manufacturing,* vol. 21, pp. 517-528, 2018.

[15] R. Chen, F. Imani, E. Reutzel and H. Yang, "From Design Complexity to Build Quality in Additive Manufacturing—A Sensor-Based Perspective," *IEEE Sensors Letters, 3(1),* pp. 1-4, 2019, .

[16] F. Froes, R. Boyer and B. Dutta, "Introduction to aerospace materials requirements and the role of additive manufacturing," *Additive Manufacturing for the Aerospace Industry,* pp. 1-6, 2019.

[17] R. Russell, D. Wells, J. Waller, B. Poorganji, E. Ott, T. Nakagawa, H. Sandoval, N. Shamsaei and M. Seifi, "Qualification and certification of metal additive manufactured hardware for aerospace applications," *Additive Manufacturing for the Aerospace Industry,* pp. 33-66, 2019.

[18] M. P. Francis, N. Kemper, Y. Maghdouri-White and N. Thayer, "Additive manufacturing for biofabricated medical device applications," *Additive Manufacturing,* pp. 311-344, 2018.

[19] J. E. Adamo, W. L. Grayson, H. Hatcher, J. S. Brown, A. Thomas, S. Hollister and S. J. Steele, "Regulatory interfaces surrounding the growing field of additive manufacturing of medical devices and biologic products," *Journal of Clinical and Translational Science,* vol. 2, no. 5, pp. 301-304, 2018.

[20] W. Huang, J. Liu, V. Chalivendra, D. Ceglarek, Z. Kong and Y. Zhou, "Statistical modal analysis for variation characterization and application in manufacturing quality control," *IIE Transactions,* vol. 46, no. 5, pp. 497-511, 2014.

[21] J. Zhong, J. Liu and J. Shi, "Predictive Control Considering Model Uncertainty for Variation Reduction in Multistage Assembly Processes," *Automation Science and Engineering, IEEE Transactions On,* vol. 7, no. 4, pp. 724-735, 2010.

[22] V. R. Joseph and C. F. Jeff Wu, "Robust Parameter Design of Multiple-Target Systems," *Technometrics,* vol. 44, no. 4, pp. 338-346, 2002.

[23] I. J. Petrick and T. W. Simpson, "3D Printing Disrupts Manufacturing: How Economies of One Create New Rules of Competition," *Research-Technology Management,* vol. 56, no. 6, pp. 12-16, 2013.

[24] S. A. Tofail, E. P. Koumoulos, A. Bandyopadhyay, S. Bose, L. O'Donoghue and C. Charitidis, "Additive manufacturing: scientific and technological challenges, market uptake and opportunities," *Materials Today,* vol. 21, no. 1, pp. 22-37, 2018.

[25] J. Shi and S. Zhou, "Quality control and improvement for multistage systems: A survey," *IIE Transactions,* vol. 41, no. 9, pp. 744, 2009.

[26] A. T. Sutton, C. S. Kriewall, M. C. Leu and J. W. Newkirk, "Powders for additive manufacturing processes: Characterization techniques and effects on part properties," in *Proceedings of the 26th Annual International Solid Freeform Fabrication Symposium–An Additive Manufacturing Conference,* 2016, .

[27] M. Grasso and B. M. Colosimo, "Process defects and in situ monitoring methods in metal powder bed fusion: a review," *Measurement Science and Technology,* vol. 28, no. 4, pp. 044005, 2017.

[28] M. Mani, S. Feng, B. Lane, A. Donmez, S. Moylan and R. Fesperman, "Measurement science needs for real-time control of additive manufacturing powder bed fusion processes," 2015.

[29] M. Mani, B. M. Lane, M. A. Donmez, S. C. Feng and S. P. Moylan, "A review on measurement science needs for real-time control of additive manufacturing metal powder bed fusion processes," *Int J Prod Res,* vol. 55, no. 5, pp. 1400-1418, 2017.

[30] S. Moylan, E. Whitenton, B. Lane and J. Slotwinski, "Infrared thermography for laser-based powder bed fusion additive manufacturing processes," in *AIP Conference Proceedings,* 2014, pp. 1191-1196.

[31] S. K. Everton, M. Hirsch, P. Stravroulakis, R. K. Leach and A. T. Clare, "Review of in-situ process monitoring and in-situ metrology for metal additive manufacturing," *Mater Des,* vol. 95, pp. 431-445, 2016.

[32] T. G. Spears and S. A. Gold, "In-process sensing in selective laser melting (SLM) additive manufacturing," *Integrating Materials and Manufacturing Innovation,* vol. 5, no. 1, pp. 16-40, 2016.

[33] G. Tapia and A. Elwany, "A review on process monitoring and control in metal-based additive manufacturing," *Journal of Manufacturing Science and Engineering,* vol. 136, no. 6, pp. 060801, 2014.

[34] W. J. Sames, F. A. List, S. Pannala, R. R. Dehoff and S. S. Babu, "The metallurgy and processing science of metal additive manufacturing," *International Materials Reviews,* vol. 61, no. 5, pp. 315-360, 2016.

[35] Y. Huang, M. C. Leu, J. Mazumder and A. Donmez, "Additive Manufacturing: Current State, Future Potential, Gaps and Needs, and Recommendations," *Transactions of the ASME, Journal of Manufacturing Science and Engineering,* vol. 137, no. 1, pp. 014001-014001-10, 2015.

[36] M. Roy, R. Yavari, C. Zhou, O. Wodo and P. Rao, "Prediction and Experimental Validation of Part Thermal History in Fused Filament Fabrication Additive Manufacturing Process," *ASME. J. Manuf. Sci. Eng.,* vol. 141, no. 12, pp. 121001, 2019.

[37] H. Wei, J. Mazumder and T. DebRoy, "Evolution of solidification texture during additive manufacturing," *Scientific Reports,* vol. 5, pp. 16446, 2015.

[38] T. DebRoy, H. L. Wei, J. S. Zuback, T. Mukherjee, J. W. Elmer, J. O. Milewski, A. M. Beese, A. Wilson-Heid, A. De and W. Zhang, "Additive manufacturing of metallic components--Process, structure and properties," *Progress in Materials Science,* vol. 92, pp. 112-224, 2018.

[39] S. A. Khairallah, A. T. Anderson, A. Rubenchik and W. E. King, "Laser powder-bed fusion additive manufacturing: Physics of complex melt flow and formation mechanisms of pores, spatter, and denudation zones," *Acta Materialia,* vol. 108, pp. 36-45, 2016.

[40] A. Dunbar, E. Denlinger, J. Heigel, P. Michaleris, P. Guerrier, R. Martukanitz and T. W. Simpson, "Development of experimental method for in situ distortion and temperature measurements during the laser powder bed fusion additive manufacturing process," *Additive Manufacturing,* vol. 12, pp. 25-30, 2016.

[41] P. Promoppatum, S. Yao, P. C. Pistorius, A. D. Rollett, P. J. Coutts, F. Lia and R. Martukanitz, "Numerical modeling and experimental validation of thermal history and microstructure for additive manufacturing of an Inconel 718 product," *Progress in Additive Manufacturing,* vol. 3, no. 1-2, pp. 15-32, 2018.

[42] H. Krauss, T. Zeugner and M. F. Zaeh, "Thermographic process monitoring in powderbed based additive manufacturing," in *AIP Conference Proceedings,* 2015, pp. 177-183.

[43] B. Lane, S. Moylan, E. P. Whitenton and L. Ma, "Thermographic measurements of the commercial laser powder bed fusion process at NIST," *Rapid Prototyping Journal,* vol. 22, no. 5, pp. 778-787, 2016.

[44] I. Gibson, D. Rosen and B. Stucker, *Additive Manufacturing Technologies Rapid Prototyping to Direct Digital Manufacturing. 2010.* Springer, .

[45] W. S. Land II, B. Zhang, J. Ziegert and A. Davies, "In-situ metrology system for laser powder bed fusion additive process," *Procedia Manufacturing,* vol. 1, pp. 393-403, 2015.

[46] P. Boulware, "Final Technical Report to National Institute of Standards and Technology and National Center for Defense Manufacturing and Machining - Measurement Science Innovation Program for Additive Manufacturing: An Evaluation of In-Process Sensing Techniques Through the Use of an Open Architecture Laser Powder Bed Fusion Platform," 2016.

[47] M. Montazeri, R. Yavari, P. Rao and P. Boulware, "In-process monitoring of material cross-contamination defects in laser powder bed fusion," *Journal of Manufacturing Science and Engineering,* vol. 140, no. 11, pp. 111001, 2018.

[48] M. Abdelrahman, E. W. Reutzel, A. R. Nassar and T. L. Starr, "Flaw detection in powder bed fusion using optical imaging," *Additive Manufacturing,* vol. 15, pp. 1-11, 2017.

[49] A. Nassar, B. Starr and E. Reutzel, "Process monitoring of directed-energy deposition of inconel-718 via plume imaging," in *Solid Freeform Fabrication Symposium (SFF),* Austin, TX, 2015, pp. 10-12.

[50] A. J. Dunbar and A. R. Nassar, "Assessment of optical emission analysis for in-process monitoring of powder bed fusion additive manufacturing," *Virtual and Physical Prototyping,* vol. 13, no. 1, pp. 14-19, 2018.

[51] A. Kramida, Y. Ralchenko and J. Reader, "NIST Atomic Spectra Database ," *National Institute of Standards and Technology (NIST),* 2017.

[52] J. C. Fox, S. P. Moylan and B. M. Lane, "Effect of Process Parameters on the Surface Roughness of Overhanging Structures in Laser Powder Bed Fusion Additive Manufacturing," *Procedia CIRP, 45*pp. 131-134, 2016, .

[53] G. Ameta, J. Fox and P. Witherell, "Tolerancing and Verification of Additive Manufactured Lattice with Supplemental Surfaces," *Procedia CIRP, 75*pp. 69-74, 2018, .

[54] Y. Lu, P. Witherell and A. Donmez, "A collaborative data management system for additive manufacturing," in *ASME 2017 International Design Engineering Technical Conferences and Computers and Information in Engineering Conference,* Quebec, Canada, 2017, pp. 1-11.

[55] C. Kan and H. Yang, "Dynamic network monitoring and control of in situ image profiles from ultraprecision machining and biomanufacturing processes," *Qual. Reliab. Eng. Int.,* vol. 33, no. 8, pp. 2003-2022, 2017.

[56] R. Chen, F. Imani, E. Reutzel and H. Yang, "From Design Complexity to Build Quality in Additive Manufacturing - A Sensor-Based Perspective," *IEEE Sensors Letters,* vol. 3, no. 1, pp. 1-4, 2018.

[57] A. Gaikwad, F. Imani, H. Yang, E. Reutzel and P. Rao, "In-situ Monitoring of Thin-Wall Build Quality in Laser Powder Bed Fusion using Deep Learning," *Smart and Sustainable Manufacturing Systems,* pp. 1-39, 2019.

[58] E. Savio, L. De Chiffre and R. Schmitt, "Metrology of freeform shaped parts," *CIRP Annals-Manufacturing Technology,* vol. 56, no. 2, pp. 810-835, 2007.

[59] C. Arrieta, S. Uribe, J. Ramos-Grez, A. Vargas, P. Irarrazaval, V. Parot and C. Tejos, "Quantitative assessments of geometric errors for rapid prototyping in medical applications," *Rapid Prototyping Journal,* vol. 18, no. 6, pp. 431-442, 2012.

[60] T. Brajlih, B. Valentan, J. Balic and I. Drstvensek, "Speed and accuracy evaluation of additive manufacturing machines," *Rapid Prototyping Journal,* vol. 17, no. 1, pp. 64-75, 2011.

[61] K. Zeng, N. Patil, H. Gu, H. Gong, D. Pal, T. Starr and B. Stucker, "Layer by layer validation of geometrical accuracy in additive manufacturing processes," in *Proceedings of the Solid Freeform Fabrication Symposium, Austin, TX, Aug,* 2013, pp. 12-14.

[62] D. Dimitrov, K. Schreve and N. De Beer, "Advances in three dimensional printing–state of the art and future perspectives," *Rapid Prototyping Journal,* vol. 12, no. 3, pp. 136-147, 2006.

[63] A. L. Cooke and J. A. Soons, "Variability in the geometric accuracy of additively manufactured test parts," in *Proceedings of 21st Annual International Solid Freeform Fabrication Symposium, Austin, TX,* 2010, pp. 1-12.

[64] J. Munguía, J. de Ciurana and C. Riba, "Pursuing successful rapid manufacturing: a users' best-practices approach," *Rapid Prototyping Journal,* vol. 14, no. 3, pp. 173-179, 2008.

[65] C. Ziemian and P. Crawn III, "Computer aided decision support for fused deposition modeling," *Rapid Prototyping Journal,* vol. 7, no. 3, pp. 138-147, 2001.

[66] A. Boschetto, V. Giordano and F. Veniali, "Surface roughness prediction in fused deposition modelling by neural networks." *Int J Adv Manuf Technol,* vol. 67, 2013.

[67] S. Bukkapatnam and B. Clark, "Dynamic modeling and monitoring of contour crafting—An extrusion-based layered manufacturing process," *Journal of Manufacturing Science and Engineering,* vol. 129, no. 1, pp. 135-142, 2007.

[68] A. Bauereiß, T. Scharowsky and C. Körner, "Defect generation and propagation mechanism during additive manufacturing by selective beam melting," *J. Mater. Process. Technol.,* vol. 214, no. 11, pp. 2522-2528, 2014.

[69] B. Yao, F. Imani, A. S. Sakpal, E. Reutzel and H. Yang, "Multifractal Analysis of Image Profiles for the Characterization and Detection of Defects in Additive Manufacturing," *ASME Journal of Manufacturing Science and Engineering,* vol. 140, no. 3, pp. 031014, 2017.

[70] F. Imani, B. Yao, R. Chen, P. Rao and H. Yang, "Joint Multifractal and Lacunarity Analysis of Image Profiles for Manufacturing Quality Control," *Journal of Manufacturing Science and Engineering,* vol. 141, no. 4, pp. 044501-044501-7, 2019.

[71] F. Imani, B. Yao, R. Chen, P. Rao and H. Yang, "Fractal pattern recognition of image profiles for manufacturing process monitoring and control," in *ASME 2018 13th International Manufacturing Science and Engineering Conference,* 2018, .

[72] F. Imani, A. Gaikwad, M. Montazeri, P. Rao, H. Yang and E. Reutzel, "Layerwise In-Process Quality Monitoring in Laser Powder Bed Fusion," *Proceedings of ASME International Manufacturing*

*Science and Engineering Conference, Volume 1: Additive Manufacturing; Bio and Sustainable Manufacturing, (51357),* pp. V001T01A038, 2018, .

[73] F. Imani, A. Gaikwad, M. Montazeri, P. Rao, H. Yang and E. Reutzel, "Process Mapping and In-Process Monitoring of Porosity in Laser Powder Bed Fusion Using Layerwise Optical Imaging," *Journal of Manufacturing Science and Engineering,* vol. 140, no. 10, pp. 101009-101014, 2018.

[74] F. Imani, R. Chen, E. Diewald, E. Reutzel and H. Yang, "Deep Learning of Variant Geometry in Layerwise Imaging Profiles for Additive Manufacturing Quality Control," *Journal of Manufacturing Science and Engineering,* vol. 141, no. 11, 2019.

[75] F. Imani, R. Chen, E. Diewald, E. W. Reutzel and H. Yang, "Image-guided variant shape analysis of layerwise build quality in additive manufacturing," in *ASME 2019 14th International Manufacturing Science and Engineering Conference,* 2019, .

[76] L. Qian, J. Mei, J. Liang and X. Wu, "Influence of position and laser power on thermal history and microstructure of direct laser fabricated Ti–6Al–4V samples," *Materials Science and Technology,* vol. 21, no. 5, pp. 597-605, 2005.

[77] E. Brandl, C. Leyens and F. Palm, "Mechanical properties of additive manufactured Ti-6Al-4V using wire and powder based processes," *Proceedings of IOP Conference Series: Materials Science and Engineering,* vol. 26, no. 1, pp. 012004, 2011.

[78] B. Roh, S. R. Kumara, T. W. Simpson and P. Witherell, "Ontology-Based Laser and Thermal Metamodels for Metal-Based Additive Manufacturing," *Proceedings of ASME 2016 International Design Engineering Technical Conferences and Computers and Information in Engineering Conference, Charlotte, North Carolina,* pp. 21-24, 2016.

[79] P. Witherell, S. Feng, T. W. Simpson, D. B. Saint John, P. Michaleris, Z. Liu, L. Chen and R. Martukanitz, "Toward Metamodels for Composable and Reusable Additive Manufacturing Process Models," *Journal of Manufacturing Science and Engineering,* vol. 136, no. 6, pp. 061025-061025-9, 2014.

[80] J. Z. Park, "Development of a Relational Energy Balance for Additive Manufacturing," *Penn State MS Thesis,* pp. 1-109, 2015.

[81] L. Hitzler, M. Merkel, W. Hall and A. Ochsner, "A Review of Metal Fabricated with Laser-and Powder-Bed Based Additive Manufacturing Techniques: Process, Nomenclature, Materials, Achievable Properties, and its Utilization in the Medical Sector," *Advanced Engineering Materials,* vol. 20, no. 5, pp. 1700658, 2018.

[82] B. H. Jared, M. A. Aguilo, L. L. Beghini, B. L. Boyce, B. W. Clark, A. Cook, B. J. Kaehr and J. Robbins, "Additive manufacturing: Toward holistic design," *Scripta Materialia,* vol. 135, pp. 141-147, 2017.

[83] D. D. Gu, W. Meiners, K. Wissenbach and R. Poprawe, "Laser additive manufacturing of metallic components: materials, processes and mechanisms," *International Materials Reviews,* vol. 57, no. 3, pp. 133-164, 2012.

[84] P. O'Regan, P. Prickett, R. Setchi, G. Hankins and N. Jones, "Metal Based Additive Layer Manufacturing: Variations, Correlations and Process Control," *Procedia Computer Science,* vol. 96, pp. 216-224, 2016.

[85] M. Samie Tootooni, A. Dsouza, R. Donovan, P. K. Rao, Z. Kong and P. Borgesen, "Classifying the Dimensional Variation in Additive Manufactured Parts From Laser-Scanned Three-Dimensional Point Cloud Data Using Maching Learning Approaches," *Journal of Manufacturing Science and Engineering,* vol. 139, no. 9, 2017.

[86] P. K. Rao, Z. Kong, C. E. Duty, R. J. Smith, V. Kunc and L. J. Love, "Assessment of Dimensional Integrity and Spatial Defect Localization in Additive Manufacturing Using Spectral Graph Theory," *ASME Journal of Manufacturing Science and Engineering,* vol. 138, no. 5, 2015.

[87] N. Boone, C. Zhu, C. Smith, I. Todd and J. R. Willmott, "Thermal near infrared monitoring system for electron beam melting with emissivity tracking," *Additive Manufacturing,* vol. 22, pp. 601-605, 2018.

[88] M. Montazeri, A. R. Nassar, C. Stutzman and P. Rao, "Heterogeneous Sensor-based Condition Monitoring in Directed Energy Deposition," *Additive Manufacturing,* vol. 30, pp. 100916, 2019.

[89] A. M. Aboutaleb, M. A. Tschopp, P. K. Rao and L. Bian, "Multi-Objective Accelerated Process Optimization of Part Geometric Accuracy in Additive Manufacturing," *ASME Journal of Manufacturing Science and Engineering,* vol. 139, no. 10, pp. 101001, 2017.

[90] V. R. Joseph, T. Dasgupta, R. Tuo and C. F. J. Wu, "Sequential Exploration of Complex Surfaces Using Minimum Energy Designs," *Technometrics,* vol. 57, no. 1, pp. 64-74, 2015.

[91] A. Bandyopadhyay and K. D. Traxel, "Metal-additive manufacturing--Modeling strategies for application-optimized designs," *Additive Manufacturing,* 2018.

[92] P. Foteinopoulos, A. Papacharalampopoulos and P. Stavropoulos, "On thermal modeling of Additive Manufacturing processes," *CIRP Journal of Manufacturing Science and Technology,* vol. 20, pp. 66-83, 2018.

[93] M. M. Francois, A. Sun, W. E. King, N. J. Henson, D. Tourret, C. A. Bronkhorst, N. N. Carlson, C. K. Newman, T. Haut, J. Bakosi, J. W. Gibbs, V. Livescu, S. A. Vander Wiel, A. J. Clarke, M. W. Schraad, T. Blacker, H. Lim, T. Rodgers, S. Owen, F. Abdeljawad, J. Madison, A. T. Anderson, J. L. Fattebert, R. M. Ferencz, N. E. Hodge, S. A. Khairallah and O. Walton, "Modeling of additive manufacturing processes for metals: Challenges and opportunities," *Current Opinion in Solid State and Materials Science,* vol. 21, no. 4, pp. 198-206, 2017.

[94] T. W. Simpson, C. B. Williams and M. and Hripko, "Preparing industry for additive manufacturing and its applications: Summary & recommendations from a National Science Foundation workshop," *Additive Manufacturing,* vol. 13, pp. 166-178, 2017.

[95] M. Markl and C. Korner, "Multiscale Modeling of Powder Bed-Based Additive Manufacturing," *Annual Review of Materials Research,* vol. 46, no. 1, pp. 93-123, 2016.

[96] T. Keller, G. Lindwall, S. Ghosh, L. Ma, B. M. Lane, F. Zhang, U. R. Kattner, E. A. Lass, J. C. Heigel, Y. Idell, M. E. Williams, A. J. Allen, J. E. Guyer and L. E. Levine, "Application of finite element, phase-field, and CALPHAD-based methods to additive manufacturing of Ni-based superalloys," *Acta Materialia,* vol. 139, pp. 244-253, 2017.

[97] M. R. Yavari, K. D. Cole and P. Rao, "Thermal Modeling in Metal Additive Manufacturing Using Graph Theory," *Journal of Manufacturing Science and Engineering,* vol. 147, no. 7, 2019.

[98] X. Wang and K. Chou, "Effect of support structures on Ti-6Al-4V overhang parts fabricated by powder bed fusion electron beam additive manufacturing," *Journal of Materials Processing Technology,* vol. 257, pp. 65-78, 2018.

[99] K. Cooper, P. Steele, B. Cheng and K. Chou, "Contact-Free Support Structures for Part Overhangs in Powder-Bed Metal Additive Manufacturing," *Inventions,* vol. 3, no. 1, 2018.

[100] H. Bikas, A. K. Lianos and P. Stavropoulos, "A design framework for additive manufacturing," *The International Journal of Advanced Manufacturing Technology,* 2019.

[101] W. King, A. Anderson, R. Ferencz, N. Hodge, C. Kamath, S. Khairallah and A. Rubenchik, "Laser powder bed fusion additive manufacturing of metals; physics, computational, and materials challenges," *Applied Physics Reviews,* vol. 2, no. 4, pp. 041304, 2015.

[102] J. Goldak, A. Chakravarti and M. Bibby, "A new finite element model for welding heat sources," *Metallurgical Transactions B,* vol. 15, no. 2, pp. 299-305, 1984.

[103] Z. Luo and Y. Zhao, "A survey of finite element analysis of temperature and thermal stress fields in powder bed fusion Additive Manufacturing," *Additive Manufacturing,* vol. 21, pp. 318-332, 2018.

[104] T. Stockman, J. A. Schneider, B. Walker and J. S. Carpenter, "A 3D Finite Difference Thermal Model Tailored for Additive Manufacturing," *Jom,* vol. 71, no. 3, pp. 1117-1126, 2019.

[105] M. R. Yavari, K. D. Cole and P. K. Rao, "Design Rules for Additive Manufacturing – Understanding the Fundamental Thermal Phenomena to Reduce Scrap," *Procedia Manufacturing,* vol. 33, pp. 375-382, 2019.

[106] D. Strong, I. Sirichakwal, G. P. Manogharan and T. Wakefield, "Current state and potential of additive–hybrid manufacturing for metal parts," *Rapid Prototyping Journal,* vol. 23, no. 3, pp. 577-588, 2017.

[107] J. Wang, X. Li and X. Zhu, "Intelligent dynamic control of stochastic economic lot scheduling by agent-based reinforcement learning," *Int J Prod Res,* vol. 50, no. 16, pp. 4381-4395, 2012.

[108] H. Meidani and R. Ghanem, "Random Markov decision processes for sustainable infrastructure systems," *Structure and Infrastructure Engineering,* pp. 1-13, 2014.

[109] B. Kazaz and T. W. Sloan, "Production policies under deteriorating process conditions," *IIE Transactions,* vol. 40, no. 3, pp. 187-205, 2008.

[110] A. H. Elwany, N. Z. Gebraeel and L. M. Maillart, "Structured Replacement Policies for Components with Complex Degradation Processes and Dedicated Sensors," *Oper. Res.,* vol. 59, no. 3, pp. 684-695, 2011.

[111] B. Yao, F. Imani and H. Yang, "Markov Decision Process for Image-Guided Additive Manufacturing," *IEEE Robotics and Automation Letters,* vol. 3, no. 4, pp. 2792-2798, 2018.

[112] B. Yao and H. Yang, "Constrained Markov Decision Process Modeling for Sequential Optimization of Additive Manufacturing Build Quality," *IEEE Access,* vol. 6, pp. 54786-54794, 2018.